\documentclass[article,english,12pt]{article}
\usepackage{amscd,amsfonts,amsmath,amssymb,amsthm,bbm,bm,latexsym,mathrsfs}
\usepackage{epsfig,graphics,graphicx,subfigure,longtable}
\usepackage[]{natbib}
\usepackage[english]{babel}
\usepackage[usenames]{color}
\usepackage{bookmark}
\usepackage{sgame}
\allowdisplaybreaks
\usepackage[top=1.2in, bottom=1.6in, left=1.3in, right=1.3in]{geometry}
\usepackage{adjustbox}
\usepackage{gensymb}
\usepackage{epstopdf}
\usepackage{tikz}
\usetikzlibrary{shapes,arrows}	
\usepackage{tkz-berge}
\usetikzlibrary{bayesnet}
\usepackage[toc,page]{appendix}
\usepackage{chapterbib}
\usepackage[normalem]{ulem}
\usepackage{cancel}

\makeatother

\theoremstyle{remark}

\theoremstyle{plain}

\newcommand{\Beta}{\mathcal{B}e}

\newcommand{\Ga}{\mathcal{G}a}

\begin{document}

\title{\vspace{-70pt}Bayesian nonparametric sparse VAR models\thanks{The authors are grateful to the Guest Editor and the reviewers for their useful comments which significantly improved the quality of the paper. We would like to thank all the conference partecipants for helpful discussions at: ``9th RCEA'' and ``12th RCEA" in Rimini; ``Internal Seminar'' at Ca' Foscari University; ``Statistics Seminar'' at University of Kent; ``9th CFE'' in London; ``ISBA 2016'' in Sardinia; ``3rd BAYSM'' at University of Florence, ``7th ESOBE'' in Venice, ``10th CFE'' in Seville, ``7th ICEEE'' in Messina, ``Bomopav 2017'' in Venice, ``Big Data in Predictive Dynamic Econometric Modeling'' at University of Pennsylvania and ``11th BNP'' in Paris. We benefited greatly from suggestions and discussions with Conception Ausin, Francis Diebold, Lorenzo Frattarolo, Pedro Galeano, Jim Griffin, Mark Jensen, Gary Koop, Dimitris Korobilis, Stefano Tonellato and Mike West.
This research used the SCSCF multiprocessor cluster system at University Ca' Foscari.}}
\author{\hspace{-20pt} Monica Billio$^{\dag}$\setcounter{footnote}{3}\footnote{Corresponding author: \href{mailto:billio@unive.it}{billio@unive.it} (M. Billio). Other contacts:  \href{mailto:r.casarin@unive.it}{r.casarin@unive.it} (R. Casarin) and \href{mailto:luca.rossini@unibz.it}{luca.rossini@unibz.it} (L. Rossini).} \hspace{15pt}
Roberto Casarin$^{\dag}$ 
\hspace{15pt}
Luca Rossini$^{\dag \ddag}$
        \\ 
        \vspace{-7pt}
        \\
        {\centering {\small{$^\dag$Ca' Foscari University of Venice, Italy \hspace{14pt}
             $^\ddag$Free University of Bozen-Bolzano, Italy}}}}             
             
\vspace{-5pt}
\date{}
\maketitle

\vspace{-10pt}

\textbf{Abstract.}  
High dimensional vector autoregressive (VAR) models require a large number of parameters to be estimated and may suffer of inferential problems. We propose a new Bayesian nonparametric (BNP) Lasso prior (BNP-Lasso) for high-dimensional VAR models that can improve estimation efficiency and prediction accuracy. Our hierarchical prior overcomes overparametrization and overfitting issues by clustering the VAR coefficients into groups and by shrinking the coefficients of each group toward a common location. Clustering and shrinking effects induced by the BNP-Lasso prior are well suited for the extraction of causal networks from time series, since they account for some stylized facts in real-world networks, which are sparsity, communities structures and heterogeneity in the edges intensity. In order to fully capture the richness of the data and to achieve a better understanding of financial and macroeconomic risk, it is therefore crucial that the model used to extract network accounts for these stylized facts. 

\medskip


\newpage

\section{Introduction}
In the last decade, high dimensional models and large datasets have increased their importance in economics \citep[e.g., see][]{ScoVar13} and finance. In macroeconomics, some authors investigate the use of large datasets (between 10 and 170 series) to improve forecasts \citep[e.g., see][]{BanGiaRe10,StoWa12,Koop13,CarClMa13,McCracken2016, KauSch12}, while in finance, large datasets (between 10 and 200 series) have been used to analyse financial crises, contagion effects and their impact on the real economy \citep[e.g., see][]{BronEngl16,BarBro16}\footnote{See Table S.1 and S.2 in Supplementary Material for a review.}. Moreover, the level of interaction, or connectedness, between financial institutions represents a powerful tool in monitoring financial stability \citep[e.g., see][]{DiebYilm15,scott2016}, whereas measuring interdependence in business cycles and in financial markets \citep{DiebYilm14, DemDieLiuYilm16} is essential in pursuing economic stability. Making inference on dependence structures in large sets of time series and representing them as graphs (or networks) becomes a relevant and challenging issue in financial and macroeconomic risk measurement \citep[e.g., see][]{BilGetLoPel12,DiebYilm15,DiebYilm16,BiaBilCasGui15}.\\
\indent For forecasting economic variables and analysing their dynamic properties and causal structure, vector autoregressive (VAR) models have been widely used. In order to avoid overparametrization and overfitting issues, Bayesian inference and suitable classes of prior distributions have been successfully used for VARs \citep[e.g., see][]{Litterman86,DoanLittermanSims84,SimsZha98}) and for panel VARs (\cite{CanovaCiccarelli})\footnote{See \cite{Kar13} for a comprehensive review.}. Nevertheless, these prior distributions may be not effective when dealing with very large VAR models. Thus, new prior distributions have been proposed to deal with zero restrictions in VAR parameters. \cite{GeoSunN08} introduce Stochastic Search Variable Selection (SSVS) based on spike-and-slab prior distribution. SSVS has been extended in many directions \citep[e.g., see][]{Koro09,KoKo15,Koro16}. \citet{wang2010sparse,AheBilCas15, BilAheCa09} propose graphical VAR relying on graphical prior distributions. \cite{Gef14} proposes Bayesian Lasso based on doubly adaptive elastic-net prior. We thus propose a new Bayesian nonparametric (BNP) prior distributions for VARs, which can improve both estimation efficiency and prediction accuracy, thanks to the combination of two types of restrictions, called clustering and shrinking effects.\\ 
\indent Our BNP Lasso prior distribution (BNP-Lasso) groups the VAR coefficients into clusters and shrinks the coefficients within a cluster toward a common location. The shrinking effects are due to the Lasso-type distribution at the first stage of the hierarchy. The Bayesian Lasso prior allows for reformulating the Bayesian VAR model as a penalized regression \citep[see][]{ParCas05} and is now a standard choice for inducing sparsity. This assumption and our inference procedure can be easily modified to combine clustering effects with alternative shrinkage procedures such as adaptive Lasso \citep{HubFel17} and adaptive elastic-net \citep{Gef14}.\\ 
\indent The clustering effects of our approach are due to a Dirichlet process (DP) hyperprior at the second stage, which induces a random partition of the parameter space. The attractive property of DP is that the number of clusters can be inferred without being specified in advance as it happens in finite mixture modelling. A similar approach for finite mixture models would be model averaging or model selection for the number of components, but it can be nontrivial to design efficient MCMC samplers (e.g. \cite{Richardson97}). Meanwhile, for DP there are relatively simple and flexible samplers for posterior approximation (e.g. \cite{walker2011}). The DP prior is standard in BNP, but the assumption could be replaced by more flexible clustering-inducing priors, such as the Pitman-Yor process prior \citep{PitmanYor1997}. In the presentation of the model we introduce two further assumptions, namely a prior partition of the parameter space and the use of conditionally independent and DP priors on each subspace. We envisage two advantages in using this specification strategy: from a modeling perspective, it adds flexibility to the model since it allows for different degrees of prior information in the blocks of parameters; from a computational perspective, it allows for breaking the posterior simulation procedure into several steps overcoming the difficulties to deal with high dimensional parameter vectors. These assumptions are not essential for the clustering and shrinking effects to appear, since the effects are given by the two-stage structure of the hyper-prior and do not depend on the blocking strategy. Moreover, the independent DP assumptions can be easily removed to allow for dependent clustering features as in beta-dependent Pitman-Yor prior \citep{Bas13,GriffinSteel2011}. We leave these issues for future research.\\ 
\indent Up to our knowledge, our paper is the first to provide sparse Bayesian nonparametric VAR models, and the proposed prior distribution  is general as it easily extends to other model classes, such as SUR models. In this sense, we contribute to the literature on Bayesian nonparametrics (\cite{Ferguson73} and \cite{Lo1984}) and its applications to time series \citep[e.g., see][]{Hir02,TaddyKottas2009,JensenMaheu2010,GriffinSteel2011,DiLucca2013,Bas13,Nieto2016,Griffin18}). See also \cite{BNP2010} for a review. We substantially improve \cite{Hir02}, \cite{Bas13} and \cite{Griffin18} by allowing for sparsity in their nonparametric dynamic models and we extend \cite{MacDun10} by proposing vectors of dependent sparse Dirichlet process priors. As regards to the posterior approximation, we develop a MCMC algorithm building on the slice sampler introduced by \cite{walker2007}, \cite{walker2011} and \cite{HatjispyrosaNicolerisaWalker} for vectors of dependent random measures.\\
\indent We also contribute to the literature on the analysis of economic and financial fluctuations through the lens of networks \citep[e.g., see][]{Acemoglu2012,acemoglu2015systemic} and the measures of risk connectedness \citep[e.g., see][]{BilGetLoPel12,DiebYilm14,DemDieLiuYilm16}. Our BNP-Lasso VAR model is particularly well suited for extracting causal networks from time series, since it allows for some features detected in many real-world networks, that are sparsity, communities or blocks, and edge heterogeneity. For the first time, we extract causal networks with clustering effects in the edge intensity. We use the posterior random partition induced by our BNP-Lasso prior to cluster the edge intensities into groups, which allows for multilayer network representation \citep{Kivela14} and thus for a better analysis of the network topology and consequently of financial and macroeconomic connectedness. In the last years, there is an emerging literature based on BNP for network analysis \citep[see, e.g.][and reference therein]{Caron2017}. Moreover, as evidenced in many papers from the network literature, edge clustering is even more important to achieve a better description of the connectivity. Our BNP-Lasso VAR allows for extracting edge clustering in a well founded inference context.\\ 
\indent We find empirical evidence of strong heterogeneity across network edges and centrality heterogeneity of nodes depending on their edge intensity levels. Finally, we also perform a comparison of our Bayesian nonparametric model with alternative shrinkage approaches and give evidence of the best forecasting performance of our proposed prior.\\ 
\indent The paper is organized as follows. Section \ref{sVAR} introduces our sparse Bayesian VAR model. Section \ref{CompDet} presents posterior approximation and network extraction methods. In Section \ref{EmpApp}, we present the applications to business cycle and realized volatility datasets. Section \ref{Concl} concludes.


\section{A sparse Bayesian VAR model}
\label{sVAR}

\subsection{VAR models}
Let $N$ be the number of units (e.g. countries, regions or micro studies) in a panel dataset and $\mathbf{y}_{i,t}= (y_{i,1t},\ldots, y_{i,mt})'$ a vector of $m$ variables available for the $i$-th unit, with $i=1,\ldots,N$. A panel VAR is defined as the system of regression equations:
\begin{equation}
\mathbf{y}_{i,t}=\mathbf{b}_i + \sum_{j=1}^N \sum_{l=1}^p B_{ijl} \mathbf{y}_{j,t-l} +  \bm{\varepsilon}_{i,t} , \quad t=1,\ldots,T\, \, \text{ and } i=1,\ldots,N, \label{VAR}
\end{equation}
where $\mathbf{b}_i=(b_{i,1},\ldots, b_{i,m})'$ the vector of constant terms and $B_{ijl}$ the $(m\times m)$ matrix of unit- and lag-specific coefficients. We assume that $\bm{\varepsilon}_{i,t}= (\varepsilon_{i,1t},\ldots, \varepsilon_{i,mt})'$, are i.i.d. for $t=1,\ldots,T$,  with Gaussian distribution $\mathcal{N}_m(\mathbf{0},\Sigma_i)$ and that $\hbox{Cov}(\bm{\varepsilon}_{i,t},\bm{\varepsilon}_{j,t})=\Sigma_{ij}$. Eq. \ref{VAR} can be written in the more compact form as
\begin{equation}
\mathbf{y}_{i,t} = B_i \mathbf{x}_t + \bm{\varepsilon}_{i,t}, \quad i=1,\ldots,N,  \label{newVAR}
\end{equation}
where $B_i = (\mathbf{b}_i,B_{i,11},\ldots,B_{i,1p},\dots,B_{i,N1},\dots,B_{i,Np})$ is a $(m\times(1+Nmp))$ matrix of coefficients and $\mathbf{x}_t = (1,\mathbf{x}_{1,t},\ldots,\mathbf{x}_{N,t})'$ is the vector of lagged variables with $\mathbf{x}_{i,t} = (y_{i,t-1},\ldots,y_{i,t-p})$ for $i=1,\ldots,N$. 

The system of equations in \eqref{newVAR} can be written in the SUR regression form: 
\begin{equation}
\mathbf{y}_t = \left(I_{Nm}\otimes \mathbf{x}_t'\right) \bm{\beta} + \bm{\varepsilon}_t, \label{NEWSUR}
\end{equation}
where $\bm{\beta}= \text{vec}(B)$, $B=(B_1', \ldots, B_N')$, $\bm{\varepsilon}_t'=(\bm{\varepsilon}_{1t}',\ldots,\bm{\varepsilon}_{Nt}')$, $\otimes$ is the Kronecker product and $\text{vec}(\cdot)$ the column-wise vectorization operator that stacks the columns of a matrix into a column vector \citep[pp.~31--32]{Magnus:99}.


\subsection{A BNP-Lasso prior assumption}
\label{PriAss}

The number of coefficients in \eqref{NEWSUR} is $n=Nm(1+Nmp)$, which can be large for high dimensional panel of time series. In order to avoid overparameterization, unstable predictions and overfitting problem, we follow a hierarchical specification strategy for the prior distributions (e.g., \cite{CanovaCiccarelli}, \cite{Kau10}, \cite{Bas13}). Some classes of hierarchical prior distributions are used to incorporate interdependences across groups of parameters with various degrees of information pooling \citep[see][for a review]{Kar13}. Other classes of priors (e.g., \cite{MacDun10}, \cite{wang2010sparse}, \cite{BiaBilCasGui15}) are used to induce sparsity. Our new class of hierarchical priors for the VAR coefficients $\bm{\beta}$ combines sparsity and information pooling within groups of coefficients. 

We assume that $\bm{\beta}$ can be exogenously partitioned in $M$ blocks, $\bm{\beta}_i = (\beta_{i1},\ldots, \beta_{in_i})$, $i=1,\ldots,M$. In our empirical applications, blocks correspond to the VAR coefficients at different lags. This assumption is not essential to appear in our prior for the sparsity and information pooling. Nevertheless, we envisage two advantages in using this specification strategy. From a modeling perspective, it adds flexibility to the model since it allows for different degrees of prior information in the blocks of parameters. For example, lag-specific shrinking effects as in a Minnesota type prior, or country-specific shrinking effects in panel VAR model can be easily introduced in our framework. From a computational perspective, it allows for breaking the posterior simulation procedure into several steps overcoming the difficulties of dealing with high dimensional parameter vectors.

For each block, we introduce shrinking effects by using a Bayesian-Lasso prior $f_i(\bm{\beta}_i)=\prod_{j=1}^{n_i} \mathcal{N}\mathcal{G}(\beta_{ij}|\mu,\gamma,\tau)$, where: 
\begin{equation}
\mathcal{N}\mathcal{G}(\beta|\mu,\gamma,\tau)= \int_0^{+\infty} \mathcal{N}(\beta| \mu,\lambda)  \Ga(\lambda|\gamma, \tau/2) d\lambda, 
\label{3}
\end{equation}
is the normal-gamma distribution  with $\mu$, $\gamma$ and $\tau$, the location, shape and scale parameter, respectively (see Appendix \ref{AppPrior}). The normal-gamma distribution induces shrinkage toward the prior mean of $\mu$. In Bayesian Lasso (\cite{ParCas05}), $\mu$ is set equal to zero. We extend the Lasso model by allowing for multiple location, shape and scale parameter such that: $f_i(\bm{\beta}_i)=\prod_{j=1}^{n_i} \mathcal{N}\mathcal{G}(\beta_{ij}|\mu_{ij}^*,\gamma_{ij}^{*},\tau_{ij}^*)$, and reduce curse of dimensionality and overfitting by employing a Dirichlet process (DP) as a prior for the normal-gamma parameters. More specifically, let $\bm{\theta}^{*}= (\mu^{*},\gamma^{*},\tau^{*})$ be the normal-gamma parameter vector, our BNP-Lasso prior for $\bm{\theta}^{*}$ and $\bm{\beta}_{i}$ is
\begin{align}
\beta_{ij}&\overset{ind}{\sim}\mathcal{N}\mathcal{G}(\beta_{ij}|	\bm{\theta}_{ij}^{*}), \quad \text{ and } \quad \bm{\theta}_{ij}^{*}|\mathbb{Q}_i\overset{i.i.d.}{\sim}\mathbb{Q}_i, \label{Hprior}
\end{align}
with $j=1,\dots,n_i$, where $\mathbb{Q}_i$ is a random measure. 

Following the strategy in \cite{Muller2004,PennellDunson2006,HatjispyrosaNicolerisaWalker}, we assume $\mathbb{Q}_i$ is a convex combination of a common $\mathbb{P}_0$ and a block-specific $\mathbb{P}_i$ random measure, that is  
\begin{align}
\mathbb{Q}_1(d\bm{\theta}_1)&=\pi_1 \mathbb{P}_0(d\bm{\theta}_1) + (1- \pi_1) \mathbb{P}_1(d\bm{\theta}_1),\\
&\vdots\nonumber\\
\mathbb{Q}_M(d\bm{\theta}_M)&=\pi_M \mathbb{P}_0(d\bm{\theta}_M) + (1- \pi_M) \mathbb{P}_M(d\bm{\theta}_M).\nonumber \label{4}
\end{align}
The random measure $\mathbb{P}_0$ favours sparsity by shrinking coefficients toward zero, as in standard Bayesian Lasso, i.e.
\begin{equation}
\mathbb{P}_{0}(d\bm{\theta})\sim \delta_{\{(0,\gamma_0,\tau_0)\}}(d(\mu,\gamma,\tau)), \quad \mbox{ with } \,\, (\gamma_0,\tau_0) \sim \mathcal{G}S(\gamma_0,\tau_0|\nu_0,p_0,s_0,n_0),
\end{equation}
where $\delta_{\{\bm{\theta}^{\ast}\}}(\bm{\theta})$ denotes the Dirac measure indicating that the random vector $\bm{\theta}$ has a degenerate distribution with mass at the location $\bm{\theta}^{\ast}$, and $\mathcal{G}S(\gamma,\tau|\nu,p,s,n)$ is a Gamma scale-shape distribution with parameters $\nu$, $p$, $s$ and $n$ (see Appendix \ref{AppPrior}). 

The random measure, $\mathbb{P}_i$, $i>0$, is a DP prior (DPP), which induces a random partition of the parameter space and a parameter clustering effect (see, \cite{Hir02}, \cite{GriffinSteel2011}, \cite{Bas13}), and shrinks coefficients toward multiple non-zero locations, i.e.
\begin{align}
\mathbb{P}_{i}(d\bm{\theta})\overset{i.i.d.}{\sim} \text{DPP}(\tilde{\alpha},H), \quad \mbox{ with } \,\,  H \sim \mathcal{N}(\mu| c,d) \cdot \mathcal{G}S(\gamma,\tau|\nu_1,p_1,s_1,n_1),
\end{align}
where $\tilde{\alpha}$ and $H$ are the DPP concentration parameter and base measure, respectively. See Appendix \ref{AppPrior} for a definition of DPP. For the mixing parameter $\pi_i$, we assume a Beta distribution, $\pi_i \overset{i.i.d.}{\sim} \Beta(\pi_i| 1,\alpha_i)$.

The amount of shrinkage in $\mathbb{P}_0$ and $\mathbb{P}_i$ is determined by the hyperparameters of $\mathcal{G}S(\nu,p,s,n)$. In our empirical application, we assume the hyperparameter values  $v_0=30$, $s_0=1/30$, $p_0=0.5$ and $n_0=18$ for the sparse component and $v_1=3$, $s_1=1/3$, $p_1=0.5$ and $n_1=10$ for the DPP component and $\alpha_i=1$ for the mixing parameter, as in \cite{MacDun10}.

For the variance-covariance matrix $\Sigma$, we assume a graphical prior distribution as in \cite{Carvalho07} and \cite{wang2010sparse}, where the zero restrictions on the covariances are induced by a graph $G$, that is by an ordered pair of sets $(N_G,E_G)$, where $N_G$ is a vertex set and $E_G$ a edge set. Conditionally to a specified graph $G$, we assume a Hyper Inverse Wishart prior distribution for $\Sigma$, that is:
\begin{equation}
\Sigma \sim \mathcal{HIW}_{G}(b,L), \label{HIWprior}
\end{equation} %
where $b$ and $L$ are the degrees of freedom and scale hyperparameters, respectively. See Appendix \ref{AppPrior} for further details on Hyper Inverse Wishart distributions. 

The prior over the graph structure is defined as a product of Bernoulli distributions with parameter $\psi$, which is the probability of having an edge. That is, a $m$-node graph $G=(N_G,E_G)$, has a prior probability:
\begin{equation}
p(G) \propto \, \prod_{i,j} \psi^{e_{ij}} (1-\psi)^{(1-e_{ij})} = \psi^{|E_G|} (1-\psi)^{\kappa-|E_G|}, \label{grap}
\end{equation} %
with $e_{ij}=1$ if $(i,j)\in E_G$, where $|N_G|$ and $|E_G|$ are the cardinalities of the vertex and edge sets, respectively,  $\kappa={|N_G| \choose 2}$ the maximum number of edges. To induce sparsity we choose $\psi=2/(p-1)$ which would provide a prior mode at $p$ edges.

\begin{figure}[t]
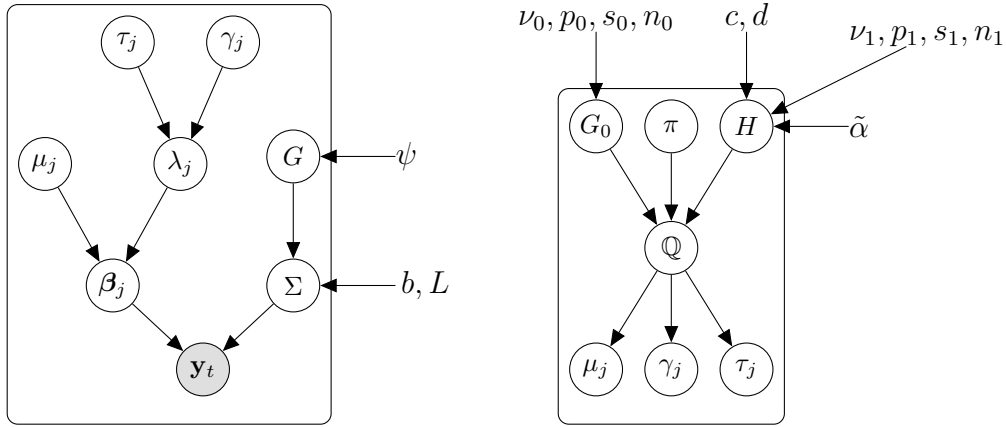

  \begin{center}
    \begin{tabular}{cc}
    \tikz{ %
        \node[obs] (y) {$\mathbf{y}_t$} ; %
        \node[latent, above=of y, xshift = -1.2cm, yshift = -0.6cm] (beta) {$\bm{\beta}_j$} ; %
        \node[latent, above=of y, xshift = 1.2cm, yshift = -0.6cm] (sigma) {$\Sigma$};
        \node[latent, above=of sigma] (G) {$G$};
        \node[latent, above=of beta, xshift = 0.9cm, yshift = -0.1cm] (lambda) {$\lambda_j$} ;
        \node[latent, above=of beta, xshift = -0.9cm, yshift = -0.1cm] (mu) {$\mu_j$} ;
          \node[latent, above=of lambda, xshift=0.7cm, yshift = -0.1cm] (gamma) {$\gamma_j$} ; %
        \node[latent, above=of lambda, xshift=-0.7cm, yshift = -0.1cm] (tau) {$\tau_j$}; %
        \node[const,  right = of sigma] (parSigma) {$\, b,L$};
        \node[const, right = of G] (parG) {$\psi$};
        \edge  {beta,sigma} {y} ; %
        \edge {parSigma} {sigma};
        \edge {parG} {G};
        \edge {G} {sigma} ;
        \edge  {lambda,mu} {beta};
        \edge {gamma,tau} {lambda};        
 \plate {yx} {(y)(beta)(sigma)(G)(lambda)(mu)(gamma)(tau)} {} ;        
           } &\hspace{10pt}
 \tikz{ %
  	\node[latent, xshift = -1cm, yshift = -0.1cm] (mu) {$\mu_j$} ;
    \node[latent, xshift=0cm, yshift = -0.1cm] (gamma) {$\gamma_j$} ; %
    \node[latent, xshift=1cm, yshift = -0.1cm] (tau) {$\tau_j$}; %
    \node[latent, above = of gamma, yshift = -0.1cm, xshift = 0cm] (hyper) {$\mathbb{Q}$};
     \node[latent, above = of hyper, yshift = -0.1cm, xshift =-1cm] (DP) {$G_0$};
     \node[latent, above = of hyper, yshift = -0.1cm, xshift = 1cm] (DP1) {$H$};
     \node[latent, above = of hyper, yshift = -0.1cm, xshift = 0cm] (pi) {$\pi$};
    \node[const, above = of DP, yshift = -0.1cm] (par) {$\nu_0,p_0,s_0,n_0$};
       \node[const, right = of DP1, yshift = 1.2cm] (parnsp) {$\nu_1,p_1,s_1,n_1$};
      \node[const, right = of DP1, yshift = 0cm] (parnsp1) {$\tilde{\alpha}$};  
      \node[const, above = of DP1, yshift = -0.1cm] (cd) {$c,d$};
       \edge {hyper} {gamma,tau,mu};
       \edge {DP,pi,DP1} {hyper};
       \edge {par} {DP};
       \edge {parnsp, parnsp1} {DP1};
       \edge {cd} {DP1};
 \plate {sec} {(mu)(gamma)(tau)(hyper)(DP)(pi)(DP1)} {} ;        
}
    \end{tabular}
  \end{center}
  \caption{DAG of the Bayesian nonparametric model for inference on VARs. It exhibits the hierarchical structure of priors and hyperparameters. The directed arrows show the causal dependence structure of the model. Left panel is related to the first stage of the hierarchy and right panel to the second stage.} \label{DAG}
\end{figure}

In summary, our hierarchical prior in Eq. \eqref{Hprior}-\eqref{grap} is represented through the \textit{Directed Acyclic Graph} (DAG) in Fig. \ref{DAG}.  Shadow and empty circles indicate the observable and non-observable random variables, respectively. The directed arrows show the causal dependence structure of the model. The left panel shows the priors for $\bm{\beta}_i$ and $\Sigma$, which are the first stage of the hierarchy.  The second stage (right panel) involves the sparse dependent DP prior for the shrinking parameters $\mu$, $\gamma$ and $\tau$. 

Our prior has the infinite mixture representation with a countably infinite number of clusters, which is one of the appealing feature of BNP:
\begin{equation}
f_i(\bm{\beta}_i|\mathbb{Q}_i) = \sum_{k=0}^\infty \check{w}_{ik} \mathcal{N}\mathcal{G}(\bm{\beta}_i|\check{\theta}_{ik}), \label{InfMix}
\end{equation}
where
\[
\check{w}_{ik}=\left\{
\begin{array}{ll}
\pi_i,&k=0,\\
(1-\pi_i)w_{ik},&k>0,\\
\end{array}
\right.
\quad
\check{\theta}_{ik}=\left\{
\begin{array}{ll}
(0,\gamma_0,\tau_0),&k=0,\\
(\mu_{ik},\gamma_{ik},\tau_{ik}),&k>0.\\
\end{array}
\right.
\]
See Appendix \ref{App_Res} for a proof. This representation shows that our prior not only shrinks coefficients to zero as in Bayesian Lasso (\cite{ParCas05}) or Elastic-net (\cite{ZouHa05}), but also induces a probabilistic clustering effect in the parameter set as in other Bayesian nonparametric models (\cite{Hir02} and \cite{Bas13}).

Our prior places no bounds on the number of mixture components and can fit arbitrary complex distribution. Nevertheless, since the mixture weights, $w_{ik}$'s, decrease exponentially quickly only a small number of clusters will be used to model the data a priori. The prior number of clusters is a random variable with distribution driven by the concentration parameter $\tilde{\alpha}$. In our applications, we set $\tilde{\alpha}=1$.

\section{Posterior inference}
\label{CompDet}

\subsection{Sampling method}
Since the posterior distribution is not tractable, Bayesian estimator cannot be obtained analytically. In this paper, we rely on simulation based inference methods, and develop a Gibbs sampler algorithm for approximating the posterior distribution. For the sake of simplicity and without loss of generality, we shall describe the sampling strategy for the two-block case, i.e. $M=2$.

In order to develop a more efficient MCMC procedure, we follow a data augmentation approach. For each block $i=1,2$, we introduce two sets of allocation variables, $\xi_{ij}, d_{ij}$, $j=1,\dots,n_i$, a set of stick-breaking variables, $v_{ij}$, $j=1,2,\ldots$ and a set of slice variables, $u_{ij}$, $j=1,\ldots,n_i$. The allocation variable, $\xi_{ij}$, selects the sparse component $\mathbb{P}_0$, when $\xi_{ij}$ is equal to zero and the non-sparse component $\mathbb{P}_i$, when it is equal to one. The second allocation variable, $d_{ij}$, selects the component of the Dirichlet mixture $\mathbb{P}_i$ to which each single coefficient $\beta_{ij}$ is allocated to. The sequence of stick-breaking variables defines the mixture weights, whereas the slice variable, $u_{ij}$, allows us to deal with the infinite number of mixture components by identifying a finite number of stick-breaking variables to be sampled and an upper bound for the allocation variables $d_{ij}$.

We demarginalize the Normal-Gamma distribution by introducing a latent variable $\lambda_{ij}$ for each $\beta_{ij}$ and obtain the joint posterior distribution
\begin{align}
&f(\Theta,\Sigma,\Lambda,U,D,V,\Xi|Y)\propto \prod_{t=1}^{T}(2\pi|\Sigma|)^{-1/2}\exp{\left(-\frac{1}{2} \left(y_t-X_t'\beta\right)' \Sigma^{-1} \left(y_t- X_t'\beta\right)\right)}\cdot \notag \\
&\,\,\prod_{j=1}^{n_1}f_1(\beta_{1j},\lambda_{1j},u_{1j},d_{1j},\xi_{1j}) \prod_{j=1}^{n_2}f_2(\beta_{2j},\lambda_{2j},u_{2j},d_{2j},\xi_{2j}) \cdot \label{JoV} \\ 
& \prod_{k>1}\Beta(v_{1k}|1,\alpha)\Beta(v_{2k}|1,\alpha)\mathcal{HIW}_{G}(b,L) \mathcal{G}S(\gamma_0,\tau_0|\nu_0,p_0,s_0,n_0) \cdot \notag \\
&\prod_{k>1}\mathcal{N}(\mu_{1k}|c,d)\mathcal{G}S(\gamma_{1k},\tau_{1k}|\nu_1,p_1,s_1,n_1) \mathcal{N}(\mu_{2k}|c,d)\mathcal{G}S(\gamma_{2k},\tau_{2k}|\nu_1,p_1,s_1,n_1),\notag
\end{align}
where $U=\{u_{ij}: j=1,2,\dots,n_i \text{ and } i=1,2\}$ and $V=\{v_{ij}: j=1,2,\dots \text{ and } i=1,2\}$ are the collections of  slice variables and stick-breaking components, respectively; $D=\{d_{ij}: j=1,2,\dots,n_i \text{ and } i=1,2\}$ and $\Xi=\{\xi_{ij}: j=1,2,\dots,n_i \text{ and } i=1,2\}$ are the allocation variables; $\Theta =\{(\mu_0,\gamma_0,\tau_0),(\mu_{ik},\gamma_{ik},\tau_{ik}): i=1,2 \text{ and } k =1,2,\dots\}$ are the atoms; $\pi=(\pi_1,\pi_2)$ are the block-specific probabilities of shrinking coefficients to zero and
\begin{align}
f_i(&\beta_{ij},\lambda_{ij},u_{ij},d_{ij},\xi_{ij})=\left(\mathbb{I}(u_{ij}<\tilde{w}_{d_{ij}})\mathcal{N}(\beta_{ij}|0,\lambda_{ij})\mathcal{G}a(\lambda_{ij}|\gamma_0,\tau_0/2)\right)^{1-\xi_{ij}}\cdot \notag\\
&\,\, \left(\mathbb{I}(u_{ij}<w_{id_{ij}}) \mathcal{N}(\beta_{ij}|\mu_{id_{ij}},\lambda_{ij})\mathcal{G}a(\lambda_{ij}|\gamma_{id_{ij}},\tau_{id_{ij}}/2)\right)^{\xi_{ij}}\pi_i^{1-\xi_{ij}}(1-\pi_i)^{\xi_{ij}}. \label{JointPos}
\end{align}
See Appendix \ref{App_Res} for a derivation.

We obtain random samples from the posterior distributions by Gibbs sampling. The Gibbs sampler iterates over the following steps using the conditional independence between variables as described in Appendix \ref{AppAFsCp}:
\begin{enumerate}
\item The slice and stick-breaking variables $U$ and $V$ are updated given $[\Theta,\bm{\beta},\Sigma,G,\Lambda,D,\Xi,\pi,Y]$;
\item The latent scale variables $\Lambda$ are updated given $[\Theta,\bm{\beta},\Sigma,G,U,V,D,\Xi,\pi,Y]$; 
\item The parameters of the Normal-Gamma distribution $\Theta$ are updated given $[\bm{\beta},\Sigma,G,\Lambda,U,V,D,\Xi,\pi,Y]$;
\item The coefficients $\bm{\beta}$ of the VAR model are updated  given $[\Theta,\Sigma,G,\Lambda,U,V,D,\Xi,\pi,Y]$;
\item The matrix of variance-covariance $\Sigma$ is updated given $[\Theta,\bm{\beta},G,\Lambda,U,V,D,\Xi,\pi,Y]$;
\item The graph G is updated given $[\Theta,\bm{\beta},\Sigma,\Lambda,U,V,D,\Xi,\pi,Y]$;
\item The allocation variables $D$ and $\Xi$ are updated given $[\Theta,\bm{\beta},\Sigma,G,\Lambda,U,V,\pi,Y]$;
\item The probability, $\pi$, of shrinking-to-zero the coefficient is updated given $[\Theta,\bm{\beta},\Sigma,G,\Lambda,U,V,D,\Xi,Y]$.
\end{enumerate}

\subsection{Network extraction}
Pairwise Granger causality has been used to extract linkages and networks describing relationships between variables of interest, e.g. financial and macroeconomic linkages \citep{BilGetLoPel12, BarBro16}. The pairwise approach does not consider conditioning on relevant covariates thus generating spurious causality effects. The conditional Granger approach includes relevant covariates, however the large number of variables relative to the number of data points can lead to overparametrization and consequently to inefficiency in gauging the causal relationships \cite[e.g., see][]{AheBilCas15, BilAheCa09}. Our BNP-Lasso can be used to extract the network while reducing overfitting and curse of dimensionality problems. Also, it allows to define edge-coloured graphs, which account for various stylized facts recently investigated in financial networks, such as the presence of communities, hubs and linkage heterogeneity.

\begin{figure}[t]
	\centering
	\tikzstyle{empty}= [circle, fill=white, draw=white, inner sep=1pt, solid,
	minimum size=2pt, font=\fontsize{10}{10}\selectfont]
	\tikzstyle{latent}= [circle, fill=black, draw=black, inner sep=1pt, solid,
	minimum size=2pt, font=\fontsize{10}{10}\selectfont]
	\tikzstyle{Edge} = [very thick]
    \setlength{\tabcolsep}{3pt}	
	\begin{tabular}{ccccc}
	(a)&(b)&(c)&(d)&(e)\\
		\begin{tikzpicture}[x=1.6cm,y=1.8cm]
		\node [latent] (v1) at (-6.5,3.4) {$$};
		\node [empty] (v1lab) at (-6.5,3.6) {$v_1$};
		\node [latent] (v2) at (-5.5,3.4) {$$};
		\node [empty] (v2lab) at (-5.5,3.6) {$v_2$};
		\node [latent] (v4) at (-6.5,2.6) {$$};
		\node [empty] (v4lab) at (-6.5,2.4) {$v_4$};
		\node [latent] (v3) at (-5.5,2.6) {$$};
		\node [empty] (v3lab) at (-5.5,2.4) {$v_3$};
		\tikzstyle{EdgeStyle}=[post]
		\tikzstyle{EdgeStyle}=[post,bend right=-40,red]
		\Edge[](v1)(v2)
		\tikzstyle{EdgeStyle}=[post,bend right=-40,blue]
		\Edge[](v1)(v3)
		\Edge[](v1)(v4)
		\Edge[](v2)(v3)
		\Edge[](v3)(v1)
		\tikzstyle{EdgeStyle}=[post,bend right=-40,red]
		\Edge[](v4)(v3)
		\Edge[](v4)(v2)
		\end{tikzpicture}&
		\begin{tikzpicture}[x=1.6cm,y=1.8cm]
		\node [latent] (v1) at (-6.5,3.4) {$$};
		\node [empty] (v1lab) at (-6.5,3.6) {$v_1$};
		\node [latent] (v2) at (-5.5,3.4) {$$};
		\node [empty] (v2lab) at (-5.5,3.6) {$v_2$};
		\node [latent] (v4) at (-6.5,2.6) {$$};
		\node [empty] (v4lab) at (-6.5,2.4) {$v_4$};
		\node [latent] (v3) at (-5.5,2.6) {$$};
		\node [empty] (v3lab) at (-5.5,2.4) {$v_3$};
		\tikzstyle{EdgeStyle}=[post]
		\tikzstyle{EdgeStyle}=[post,bend right=-40,red]
		\Edge[](v1)(v2)
		\tikzstyle{EdgeStyle}=[post,bend right=-40,red]
		\Edge[](v4)(v3)
		\Edge[](v4)(v2)
		\end{tikzpicture}&
	     \begin{tikzpicture}[x=1.6cm,y=1.8cm]
		\node [latent] (v1) at (-6.5,3.4) {$$};
		\node [empty] (v1lab) at (-6.5,3.6) {$v_1$};
		\node [latent] (v2) at (-5.5,3.4) {$$};
		\node [empty] (v2lab) at (-5.5,3.6) {$v_2$};
		\node [latent] (v4) at (-6.5,2.6) {$$};
		\node [empty] (v4lab) at (-6.5,2.4) {$v_4$};
		\node [latent] (v3) at (-5.5,2.6) {$$};
		\node [empty] (v3lab) at (-5.5,2.4) {$v_3$};
		\tikzstyle{EdgeStyle}=[post]
		\tikzstyle{EdgeStyle}=[post,bend right=-40,blue]
		\Edge[](v1)(v3)
		\Edge[](v1)(v4)
		\Edge[](v2)(v3)
		\Edge[](v3)(v1)
		%
		\end{tikzpicture}&
		\begin{tikzpicture}[x=1.6cm,y=1.8cm]
		\node [latent] (v1) at (-6.5,3.4) {$$};
		\node [empty] (v1lab) at (-6.5,3.6) {$v_1$};
		\node [latent] (v2) at (-5.5,3.4) {$$};
		\node [empty] (v2lab) at (-5.5,3.6) {$v_2$};
		\node [latent] (v4) at (-6.5,2.6) {$$};
		\node [empty] (v4lab) at (-6.5,2.4) {$v_4$};
		\node [latent] (v3) at (-5.5,2.6) {$$};
		\node [empty] (v3lab) at (-5.5,2.4) {$v_3$};
		\tikzstyle{EdgeStyle}=[post]
		\tikzstyle{EdgeStyle}=[post,bend right=-40,red]
		\Edge[](v1)(v2)
		\Edge[](v2)(v1)
		\tikzstyle{EdgeStyle}=[post,bend right=-40,blue]
		\Edge[](v4)(v3)
		\Edge[](v3)(v4)
		%
		\end{tikzpicture}&
		\begin{tikzpicture}[x=1.6cm,y=1.8cm]
		\node [latent] (v1) at (-6.5,3.4) {$$};
		\node [empty] (v1lab) at (-6.5,3.6) {$v_1$};
		\node [latent] (v2) at (-5.5,3.4) {$$};
		\node [empty] (v2lab) at (-5.5,3.6) {$v_2$};
		\node [latent] (v4) at (-6.5,2.6) {$$};
		\node [empty] (v4lab) at (-6.5,2.4) {$v_4$};
		\node [latent] (v3) at (-5.5,2.6) {$$};
		\node [empty] (v3lab) at (-5.5,2.4) {$v_3$};
		\tikzstyle{EdgeStyle}=[post]
		\tikzstyle{EdgeStyle}=[post,bend right=-40,red]
		\Edge[](v1)(v2)
		\Edge[](v1)(v3)
		\tikzstyle{EdgeStyle}=[post,bend right=-40,blue]
		\Edge[](v1)(v4)
		%
		\end{tikzpicture}
		\\
		\vspace{2pt}
		\begin{scriptsize}		
		$
		        \begingroup 
         \setlength\arraycolsep{5pt}		
         \begin{pmatrix}  
		0 & 0 & 1 & 0\\
		1 &	0 &	0 &	1\\
		1 &	1 &	0 &	1\\
		1 &	0 &	0 &	0\\
		\end{pmatrix}
		\endgroup
		$
		\end{scriptsize}
		& 
		\vspace{2pt}
		\begin{scriptsize}		
		$\begin{pmatrix} 
		0 & 0 & 0 & 0\\
		1 &	0 &	0 &	1\\
		0 &	0 &	0 &	1\\
		0 &	0 &	0 &	0\\
		\end{pmatrix}$
		\end{scriptsize}
		&
		\vspace{2pt}
		\begin{scriptsize}		
		$\begin{pmatrix} 
		0 & 0 & 1 & 0\\
		0 &	0 &	0 &	0\\
		1 &	1 &	0 &	0\\
		1 &	0 &	0 &	0\\
		\end{pmatrix}$
		\end{scriptsize}
		&
		\vspace{2pt}
		\begin{scriptsize}		
		$\begin{pmatrix} 
		0 & 1 & 0 & 0\\
		1 &	0 &	0 &	0\\
		0 &	0 &	0 &	1\\
		0 &	0 &	1 &	0\\
		\end{pmatrix}$
		\end{scriptsize}
		&
		\vspace{2pt}
		\begin{scriptsize}		
		$\begin{pmatrix} 
		0 & 0 & 0 & 0\\
		1 &	0 &	0 &	0\\
		1 &	0 &	0 &	0\\
		1 &	0 &	0 &	0\\
		\end{pmatrix}$
		\end{scriptsize}
		\\
		\vspace{2pt}
		\begin{scriptsize}
		$
        \begingroup 
         \setlength\arraycolsep{2pt}		
         \begin{pmatrix} 
		0 & 0 & 0.1 & 0\\
		0.3 &	0 &	0 &	0.3\\
		0.1 &	0.1 &	0 &	0.3\\
		0.1 &	0 &	0 &	0\\
		\end{pmatrix}
		\endgroup
		$
         \end{scriptsize}&
         		\vspace{2pt}
		\begin{scriptsize}
         $        
         \begingroup 
         \setlength\arraycolsep{4pt}		
         \begin{pmatrix} 
		0 & 0 & 0 & 0\\
		0.3 &	0 &	0 &	0.3\\
		0 &	0 &	0 &	0.3\\
		0 &	0 &	0 &	0\\
		\end{pmatrix}
		\endgroup
		$
         \end{scriptsize}&
         		\vspace{2pt}
		\begin{scriptsize}
		$
		\begingroup 
         \setlength\arraycolsep{2pt}		
         \begin{pmatrix} 
		0 & 0 & 0.1 & 0\\
		0 &	0 &	0 &	0\\
		0.1 &	0.1 &	0 &	0\\
		0.1 &	0 &	0 &	0\\
		\end{pmatrix}
		\endgroup
		$
         \end{scriptsize}
         &
         \vspace{2pt}
		\begin{scriptsize}
		$
        \begingroup 
         \setlength\arraycolsep{2pt}		
         \begin{pmatrix} 
		0 & 0.3 & 0 & 0\\
		0.3 &	0 &	0 &	0\\
		0 &	0 &	0  &	0.1\\
		0 &	0 &	0.1 &0\\
		\end{pmatrix}
		\endgroup
		$
         \end{scriptsize}
         &
         \vspace{2pt}
		\begin{scriptsize}
		$
		        \begingroup 
         \setlength\arraycolsep{4.5pt}		
         \begin{pmatrix}  
		0 &   0 & 0 & 0\\
		0.3 &	0 &	0 &	0\\
		0.3 &	0 &	0  &0	\\
		0.1 &	0 &	0 &0\\
		\end{pmatrix}
		\endgroup
		$
         \end{scriptsize}\\
	\end{tabular}
	\caption{
Weighted graphs with $\hat{K}=2$ intensity levels: $\hat{\mu}_{1}^{\ast}=0.1$ (blue edges) and $\hat{\mu}_{2}^{\ast}=0.3$ (red edges). In each graph the node $v_i$ represents the variable $i$ in the 4-dimensional VAR(1), a clockwise-oriented edge from node $j$ to node $i$ represents a non-null coefficient for the variable $y_{j,t-1}$ in the $i$-th equation of the VAR. Panel (a): weighted graph $G=(V,E,C)$ (top) with vertex set $V=\{v_1,v_2,v_3,v_4\}$, edge set $E=\{e_1,e_2,e_3,e_4\}$, where $e_1=\{v_1,v_2\}$, $e_2=\{v_1,v_3\}$, $e_3=\{v_1,v_4\}$, $e_4=\{v_2,v_3\}$, $e_5=\{v_3,v_1\}$, $e_6=\{v_4,v_2\}$, $e_7=\{v_4,v_3\}$ and the adjacency matrix $A$ (middle) and the weights matrix $C$ (bottom).  Panel (b): the subgraph $G_1=(V,E_1)$ with $E_1=\{e_1,e_6,e_7\}$ induced by edges of intensity level $\hat{\mu}_{1}^{\ast}=0.3$. Panel (c): the subgraph $G_2=(V,E_2)$ with $E_2=\{e_2,e_3,e_4,e_5\}$ induced by edges with intensity $\hat{\mu}_{2}^{\ast}=0.3$. Panel (d): graph with two communities (red $V_1=\{v_1,v_2\}$ and blue $V_{2}=\{v_3,v_4\}$). Panel (e): graph with a hub (node $v_1$).}
	\label{fig:graph_ex}
\end{figure}
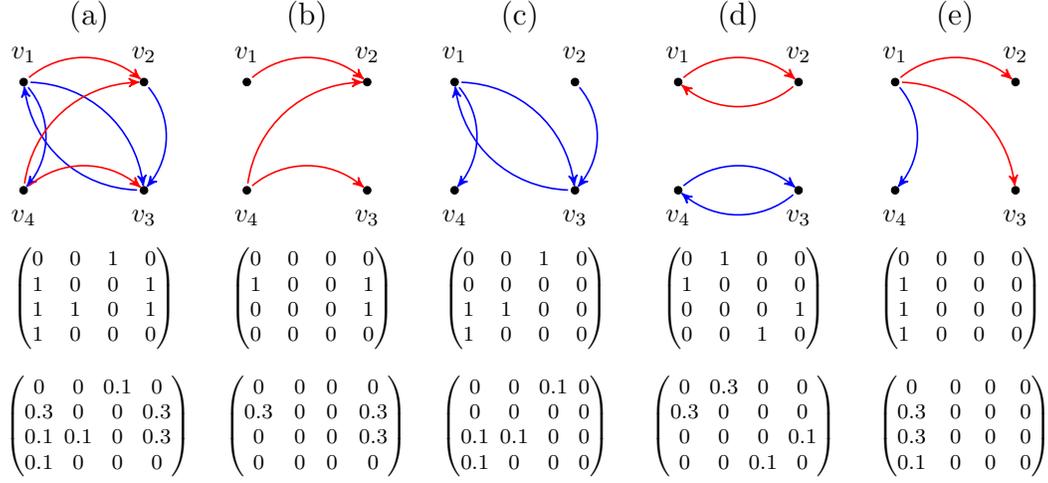

For sake of simplicity, we consider one lag, one block of coefficients and one unit, i.e. $M=1$, $N=1$ and $p=1$. We denote with $G =(V,E)$ the directed graph with vertex set $V=\{1,\ldots,m\}$ and edge set $E\subset V\times V$. A directed edge from $i$ to $j$ is the ordered set $\{i,j\} \in E$ with $i,j\in V$. The adjacency matrix $A$ associated with $G$ has $(j,i)$-th element
\[
a_{j,i}=\left\{
\begin{array}{ll}
1& \text{ if } \{i,j\} \in E,\\
0& \text{ otherwise}.\\
\end{array}
\right.
\]
See \cite{newman2010} for an introduction to graph theory and network analysis and \cite{Jackson:2008} for an economic perspective on networks. 
A weighted graph is defined by the ordered triplet $G=(V,E,C)$, where $C$ is the weight matrix with $(j,i)$-th element $c_{i,j}$ representing the weight associated to the edge $\{i,j\}$. When $c_{i,j}$ belongs to a countable set, then $G=(V,E,C)$ is called edge coloured graph.\\
\indent In time series graphs, causal networks encode the conditional independence structure of a multivariate process \citep{Eichler2013}. An edge $\{i,j\} \in E$ exists if and only if the VAR coefficient $B_{ji}$ of the variable $y_{i,t-1}$ in the equation of $y_{j,t}$ is not null. An advantage in using BNP-Lasso VAR is that the inference procedure provides edge probabilities, edge weights and edge partition as a natural output and graph uncertainty can be easily included in network analysis. 
More specifically, we assume an edge from $i$ to $j$ exists, i.e. $\{i,j\}\in E$, if $\hat{\xi}_{1,g(i,j)} = 0$,  where $\hat{\xi}_{1,g(i,j)}$ is the Maximum a Posterior (MAP) estimator of $\xi_{1,g(i,j)}$. The posterior clustering of the VAR coefficients allows us to define an edge coloured graph $G=(V,E,C)$. Following a standard procedure in BNP \citep[e.g., see][]{Bas13}, we use the MCMC draws of the allocation variables $d_{1i}$ and $d_{1j}$ to approximate the posterior probabilities of joint classification of an edge $\phi_{ij}=P(d_{1i}=d_{1j}|Y,\xi_{1i}=0,\xi_{1j}=0)$, 
\begin{equation}
\hat{\phi}_{ij}=\frac{1}{H} \sum_{h=1}^{H} \delta_{d_{1i}^h}(d_{1j}^h) \label{Pair_Pos_Matr}
\end{equation}
and to find 
the posterior partition $\hat{D}$, which minimizes the sum of squared deviations from the joint classification probabilities, i.e.
\begin{equation}
\hat{D}= \underset{D \in \{D^1,\ldots, D^H\}}{\arg \min} \sum_{i=1}^{\tilde{m}} \sum_{j=1}^{\tilde{m}} \left(\delta_{d_{1i}^h}(d_{1j}^h)-\hat{\phi}_{ij}\right)^2. \label{Pair_Post_Pb}
\end{equation} 
The number of distinct elements $\hat{K}$ in the partition $\hat{D}$ provides an estimate of the number of edge intensity levels (colours). The weight $c_{ji}$  indicating  the intensity of the edge $\{i,j\}$ can be computed by using the location atom posterior mean, $\hat{\mu}^{\ast}_{kl}$,
\begin{equation}
c_{ji} = \left\{
\begin{array}{ll}
\hat{\mu}^{\ast}_{11} & \text{ if } \hat{d}_{1,g(i,j)} = 1 \mbox{ and } \hat{\xi}_{g(i,j),l}=0, \\
\vdots & \vdots \\
\hat{\mu}^{\ast}_{1\hat{K}} & \text{ if } \hat{d}_{1,g(i,j)} = \hat{K} \mbox{ and } \hat{\xi}_{1,g(i,j)}=0, \\
0 & \text{ if } \hat{\xi}_{1,g(i,j)}=1. \\
\end{array}
\right. \label{Eq_Color}
\end{equation}
Panel (a) in Fig. \ref{fig:graph_ex} provides an example of weighted network representation for a VAR(1) with $m=4$ variables and $\hat{K}=2$ intensity levels.

A better understanding of the nodes centrality and of the risk connectedness can be achieved by exploiting the information in the edge weights \cite[e.g., see][]{BiaBilCasGui15}. One can view the number of edges as more important than the weights, so that the presence of many edges with any weight might be considered more important than the total sum of edge weights. However, edges with large weights might be considered to have a much greater impact than edges with only small weights \citep[e.g., see][]{OPSAHL2010245}. BNP-Lasso VAR can be used to better analyse the contribution of the weights to nodes centrality. The posterior partition of the edges (VAR coefficients) into a finite number of groups allows us to represent $G=(V,E,C)$ as a $\hat{K}$-multilayer graph $G=(V,\{E_k\}_{k=1}^{\hat{K}},\{C_k\}_{k=1}^{\hat{K}})$, where the $k$-th layer is the sub-graph $G_k=(V,E_k)$, with the set $E_k=\{\{i,j\}\in E; c_{ij}=\hat{\mu}^{\ast}_{k}\}$ of all edges with intensity $\hat{\mu}^{\ast}_{k}$ (see \cite{Kivela14}). Panels (b) and (c) of Fig. \ref{fig:graph_ex} provide an example with two layers.\\ 
\indent Node centrality and network connectivity can be now analysed exploiting the multilayer representation. The out-degree, $\omega_{i}^{+}$, and weighted out-degree \citep{OPSAHL2010245}, $\sigma_{i}^{+}$, of the node $i$ can be decomposed along the $\hat{K}$  layers
\begin{eqnarray}\label{dec}
\omega_{i}^{+} = \sum_{j=1}^m a_{ji} = \sum_{k=1}^{\hat{K}} \omega_{i,k}^{+}, \quad \text{where} \,\, \omega_{i,k}^{+} = \sum_{j=1}^m a_{ji} \mathbb{I}(c_{ij}=\hat{\mu}^{\ast}_{k}),\\
\sigma_{i}^{+} = \sum_{j=1}^m c_{ji} = \sum_{k=1}^{\hat{K}} \sigma_{i,k}^{+} = \sum_{k=1}^{\hat{K}} \hat{\mu}^{\ast}_{k} \omega_{i,k}^{+}, \quad \text{where} \,\, \sigma_{i,k}^{+} = \sum_{j=1}^m c_{ji} \mathbb{I}(c_{ij}=\hat{\mu}^{\ast}_{k}).
\end{eqnarray}
Similarly, it is possible to decompose the vertex in-degree and total-degree measures. These decompositions are useful in quantifying the contribution of each node to the system-wide connectedness. In our financial and macroeconomic applications, we find that nodes with large out-degree become less relevant when edge intensity is considered. To exemplify, node $v_1$ has the largest out-degree, $\omega_{1}^{+}=3$, in Panel (a), Fig. \ref{fig:graph_ex}, but is not central in the layer $G_1$ (Panel (b)). In this layer node $v_1$ has out-degree $\omega_{1,2}^{+}=1$ whereas node $v_4$ has the largest out-degree $\omega_{4,2}^{+}=2$. Following the weighted degree measure, node $v_4$ is central in $G$ and in $G_1$ with out-degree, $\sigma_{4}^{+}=0.6$ and $\sigma_{4,1}^{+}=0.6$, respectively, but it is not central in $G_2$, where its degree is $\sigma_{4,2}^{+}=0$.\\
 \indent The connectivity analysis presented above relies on the number of direct connections among nodes. In order to gain a broader idea of the network connectivity patterns, indirect connections need to be considered. The notion of path $s_{ij}$ between two vertices $i$ and $j$, called endvertices, accounts for the indirect connections between them. A path $s_{ij}$ is defined by ordered sequences of distinct vertices $V(s_{ij})=\{v_0,v_1,\ldots,v_l\}$ and edges $E(s_{ij})=\{e_1,\ldots,e_{l}\}\subset V$, with $e_1=(v_0,v_1)$, $e_{l}=\{v_{l-1},v_{l}\}$, and $v_0=i$ and $v_{l}=j$. The number of edges $|E(s_{ij})|=l$ in a path is called path length. \\ 
 \indent The distance between two nodes $i$ and $j$ is defined as the length of the shortest path, i.e. $d(i,j)=\min\{|E(s_{ij})|,\, \hbox{s.t.}\, s_{ij}\, \hbox{is a path between}\, i\, \hbox{and}\, j\}$. It can be used to define the $d$-step ego network of a node $i\in E$, that is a sub-graph of $G$ induced by $i$ and all nodes with distance less than $d$ from $i$, $V(i)=\{j\in E,\, \hbox{s.t.}\, d(i,j)\leq d\}$ and $E(i)=\{\{i,j\}\in E,\, \hbox{s.t.}\,i,j\in V(i)\}\cup \{\{j,i\}\in E,\, \hbox{s.t.}\,i,j\in V(i)\} $. In our empirical applications, ego-networks will be used to show the heterogeneity in the linkages strength of a given node. \\
\indent The distance is a local connectivity measure, which does not account for the whole interconnected topology, thus we consider the average path length (APL), that is a global connectivity measure defined as
\begin{equation}
APL(G)=\frac{1}{m(m-1)}\sum_{i=1}^{m}\sum_{j=1}^{m}d(i,j).
\end{equation}
A low APL may indicate a random graph structure of either  Erd\"{o}s-R\'{e}nyi type or  Watts-Strogatz type (see \cite{newman2010}). APL assumes shocks follow the shortest possible path, but  economic shocks are in general unlikely to be restricted to follow specific paths and are also likely to have feedback effects.\\
\indent A measure which accounts for all possible paths connecting two nodes is the eigenvector centrality (modified Bonacich's beta centrality, see \cite{BiaBilCasGui15})
\begin{equation}
h(G) = \left(I_m - \frac{1}{(m-1)^2} AA' \right)^{-1} \left(I_m - \frac{1}{(m-1)} A \right)\bm{\iota}, \,\, \text{with } \bm{\iota} = (1,\ldots, 1)'.
\end{equation}
If a VAR of order $p>1$ is used, then not only direct effects from $y_{i,t-p}$ to $y_{j,t}$ should be considered as in Granger causality,  but also indirect causal effects from $y_{i,t-p}$ to $y_{j,t}$ through $y_{k,t-l}$, $1<l<p$, as in Sims causality \citep{Eichler2013}. Moreover, VAR shocks can be correlated and their identification can be challenging in high dimension. In order to solve all these issues, \cite{DiebYilm14} propose a global connectivity measure, which relies on a generalized identification framework and on the $H$-step ahead forecast error variance
\begin{equation}
\Theta(G) = \sum_{i,j = 1, i\ne j}^m \sum_{h=0}^{H-1} \frac{\left(\mathbf{e}_i'\Phi_h \Sigma \mathbf{e}_j\right)^2}{m \sigma_{jj}c_i}, \, \, H=1,2,\ldots,   \label{Diebeq}
\end{equation} 
where $c_i$ is a normalizing constant and $\Phi_h$ satisfies the recursion $\Phi_h = B_1\Phi_{h-1}+B_2\Phi_{h-2}+ \ldots + B_p \Phi_{h-p}$ with $\Phi_0 = I_m$ and $\Phi_h = O_m$ for $h<0$.  The second and third term of the decomposition 
\begin{equation}
\left(\mathbf{e}_i'\Phi_h \Sigma \mathbf{e}_j\right)^2 = \left( \sum_{l=1}^{h-1} \mathbf{e}_i' B_l\Phi_{h-l} \Sigma \mathbf{e}_j \right)^2 \!+ 2\left( \sum_{l=1}^{h-1} \mathbf{e}_i' B_l\Phi_{h-l} \Sigma \mathbf{e}_j \right)\left(\mathbf{e}_i'B_h \Sigma \mathbf{e}_j \right) + \left(\mathbf{e}_i' B_h \Sigma \mathbf{e}_j\right)^2 \label{Decom}
\end{equation}
provide the contribution of the $h$-lag connectivity structure to the global measure. See Appendix \ref{App_Res} for a derivation.
The presence of nodes with large out-degree in the $h$-th lag graph can impact significantly on the system-wide connectedness.  Both global measures will be used in combination with multilayer decomposition of the coloured graph.\\ 
\indent Finally, other two useful notions are the ones of community and hub. Assume the vertex set $V$ is partitioned in a sequence of disjoint subsets $V_1,\ldots, V_K$, then an element $V_k$ of the partition is called community. An example of community structured graph is in Panel (d) of Fig. \ref{fig:graph_ex}. By construction, the partition of the nodes induces an edges partition (e.g., communities $V_1$ and $V_2$ in Panel (d)). Since our BNP-Lasso prior allows for any type of partition of the edge set, then typical community structures appearing in stochastic block models and modular graphs \citep[see][]{newman2010} can be extracted with our model. Panel (e) of Fig. \ref{fig:graph_ex} provides an example of hub, that is a node with a larger number of edges in comparison with other nodes in the graph. Node $v_1$ is an hub in the graph of Panel (e), since it has out-degree 3 whereas the other nodes have null out-degree.

\subsection{Simulation experiments} \label{Sim}
We study the goodness of fit of our model and, following the standard practice in BNP analysis (e.g., see \cite{Griffin_06}, \cite{GriffinSteel2011}, \cite{Griffin18}), we simulate data from a parametric model, which is in the family of the likelihood of our model. We consider a VAR(1) model
\begin{equation}
\textbf{y}_t= B\textbf{y}_{t-1} + \bm{\varepsilon}_t, \quad \bm{\varepsilon}_t \overset{i.i.d.}{\sim} \mathcal{N}_m(\bm{0},\Sigma) \quad t=1,\dots, 100, \notag
\end{equation}
where the matrix $B$ represents the collection of pairwise relationships between the time series in the panel (i.e., countries or institutions). A non-zero coefficient indicates there is a linkage between two nodes in the network.\\
\indent We consider three dimension settings for $\mathbf{y}_t$ ($m=20$ (small), $m=40$ (medium) and $m=80$ (large)) and different parameter settings. For each setting, we conduct $50$ experiments, generating $50$ independent datasets and estimating our BNP-Lasso VAR model. In all the experiments, we have chosen the hyperparameters as in Section \ref{PriAss}, and set the degree of freedom parameter, $b_0=3$ and the scale matrix, $L=I_n$. We iterate $5,000$ times the Gibbs sampler described in Section \ref{CompDet}, and after discarding a burn-in sample and applying thinning, we use the MCMC samples of $\Xi$ to estimate the network adjacency matrix and the samples of $D$ to estimate the number of colours and classify the edge intensity. We provide in Fig. \ref{graphXi} the typical causality network estimated in four experiments. See Supplementary Material for further results.
\begin{figure}[t]
 \centering
 \begin{tabular}{cc}
   \includegraphics[height=4cm, width=5.9cm]{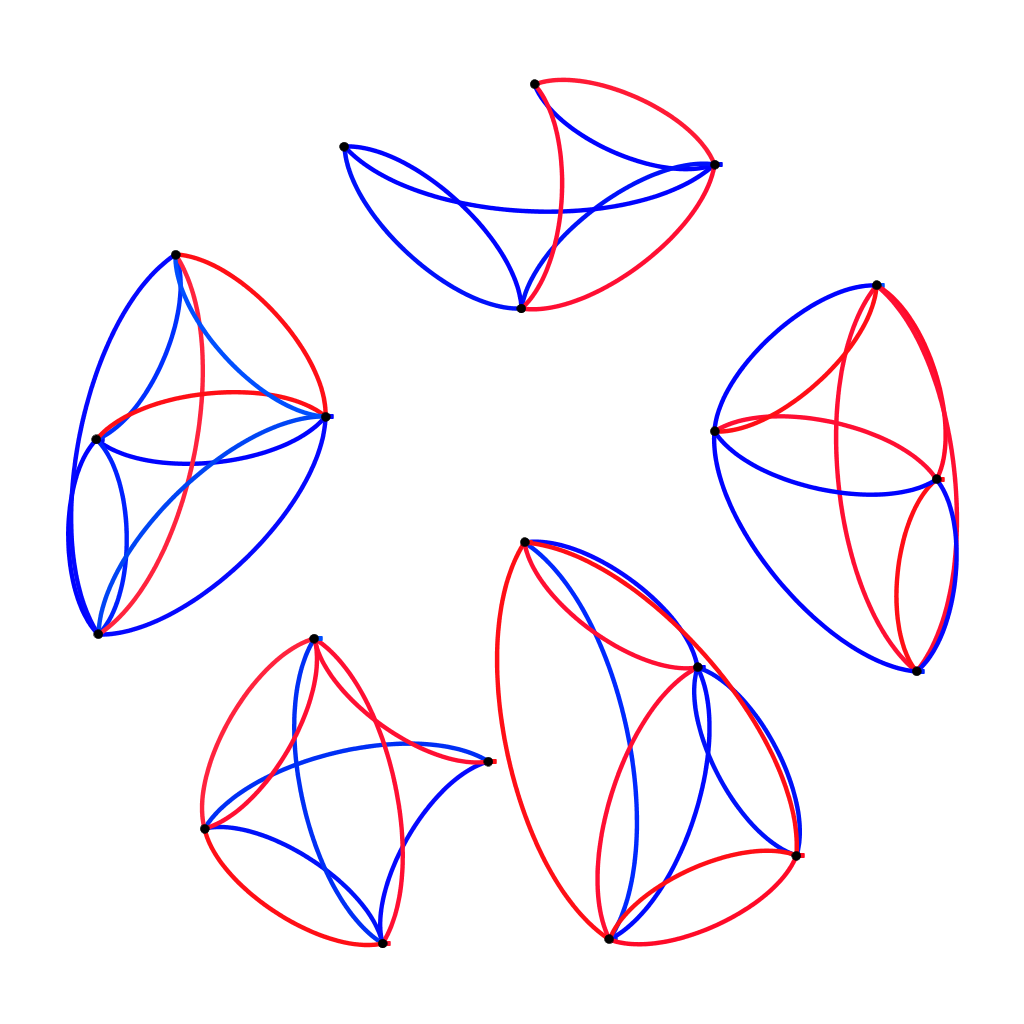} &
     \includegraphics[height=4cm, width=5.9cm]{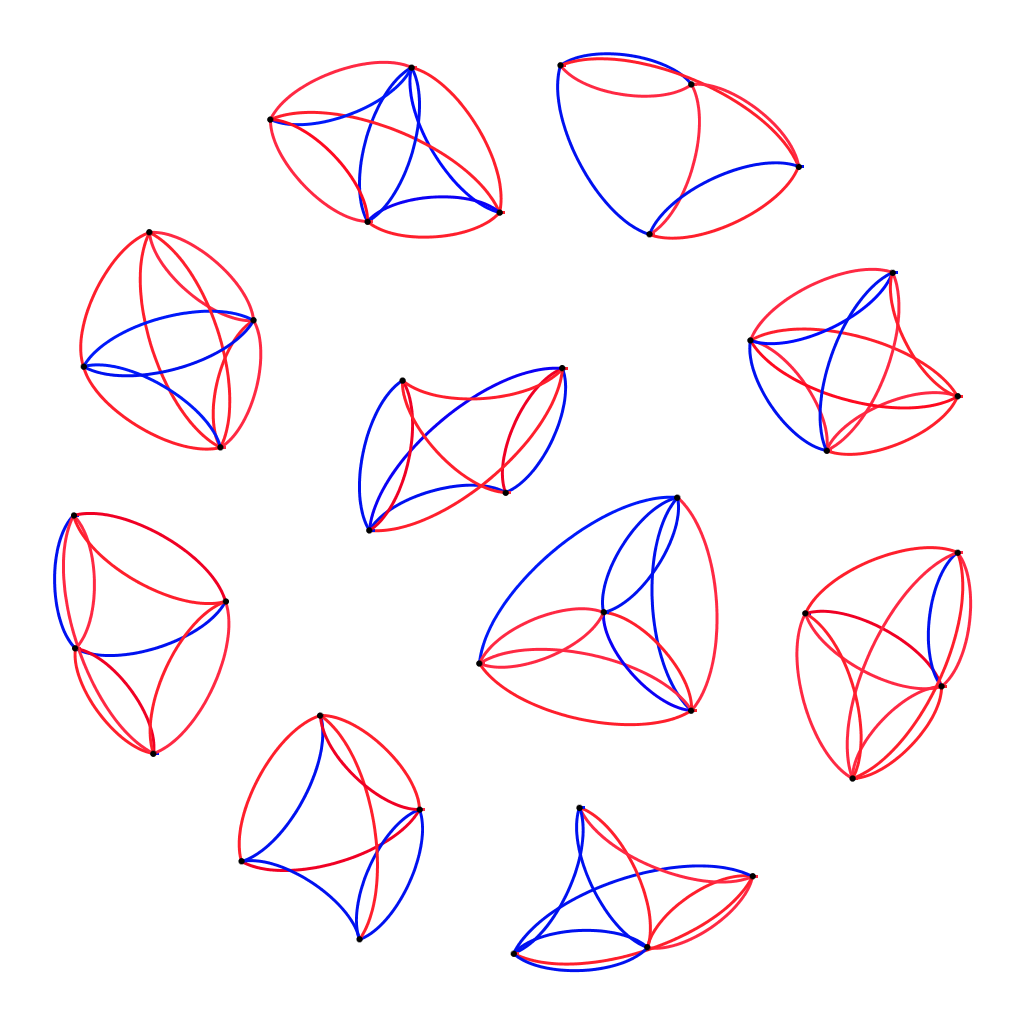} \\
        {\small (a) $m=20$} & (b) {\small $m=40$} \\
   \includegraphics[height=4cm, width=5.9cm]{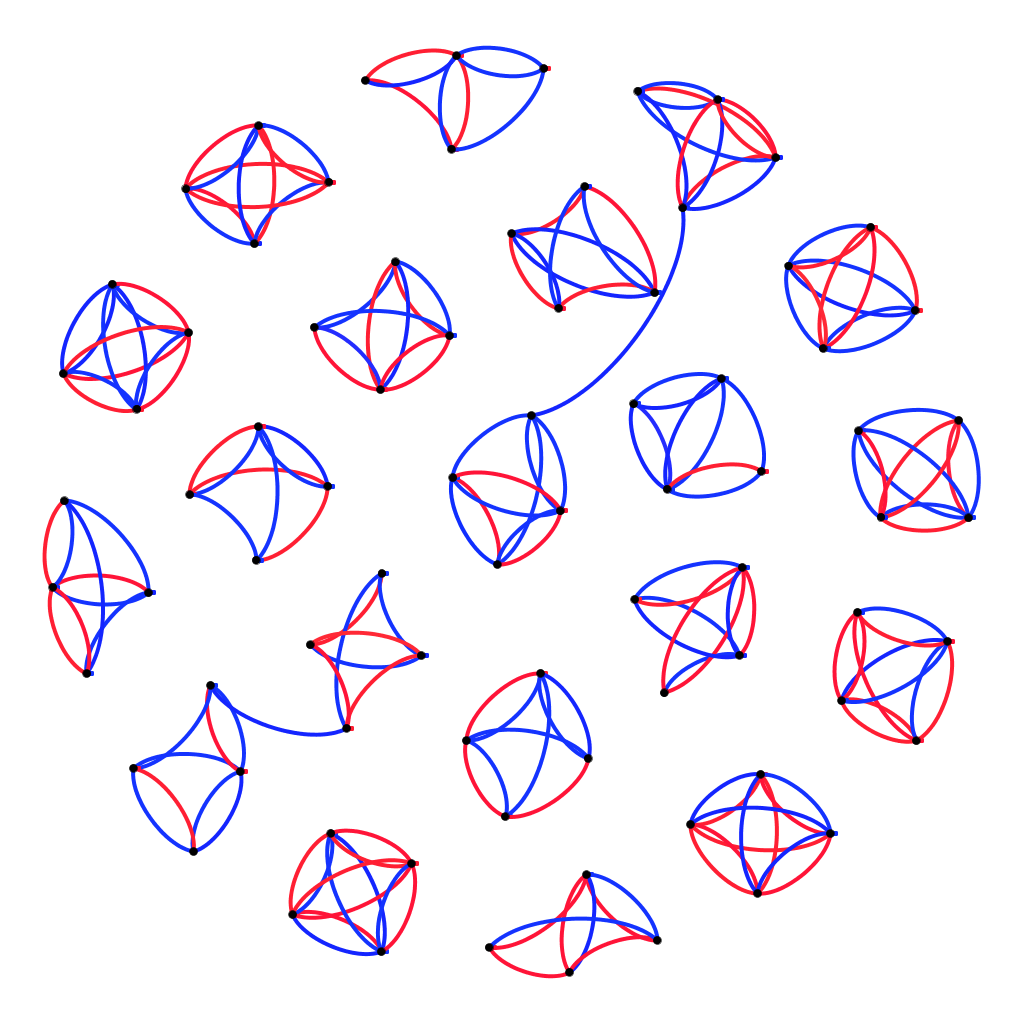} & 
 \includegraphics[height=4cm, width=5.9cm]{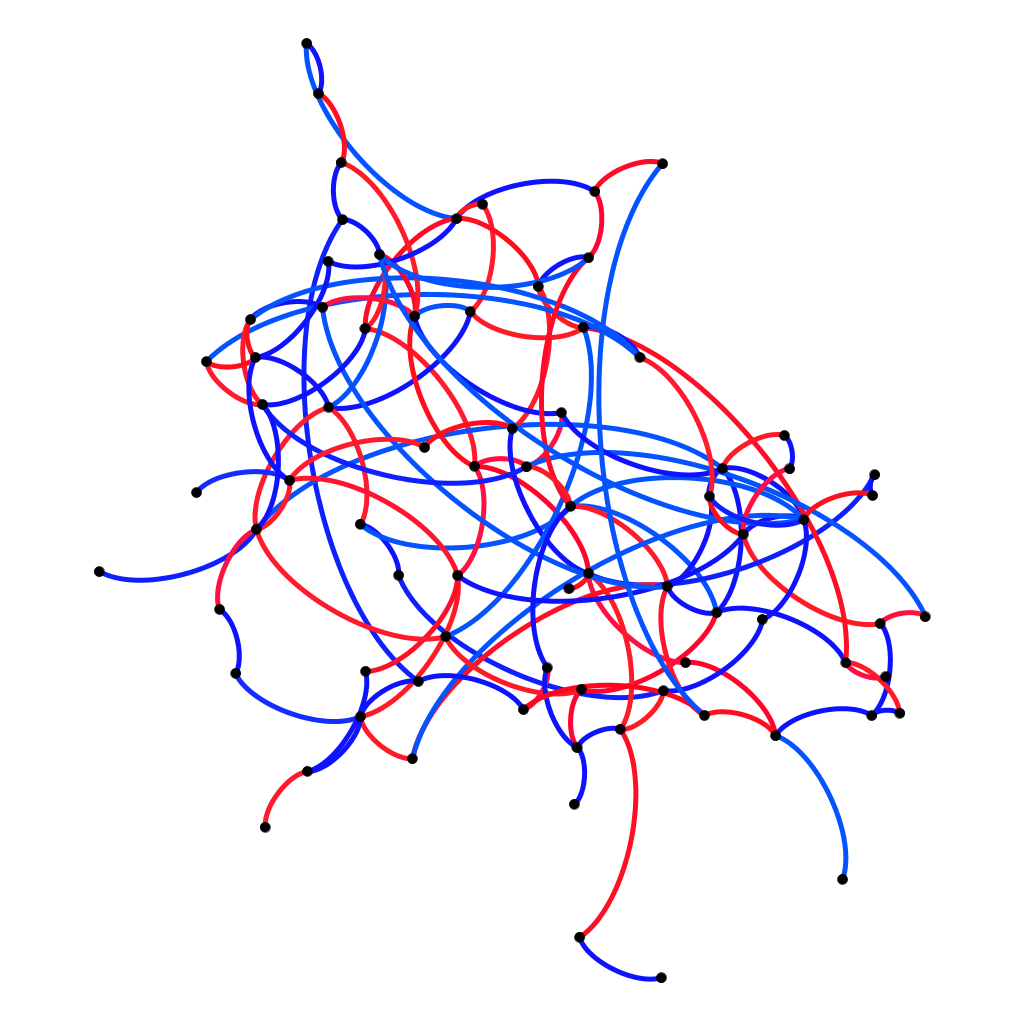} \\
  {\small (c) $m=80$ with block entries} & (d) {\small $m=80$ with random entries} \\
    \end{tabular}
 \caption{Weighted network estimated in four Monte Carlo experiments, for different model dimensions $m=20$ (panel (a)), $m=40$ (panel (b)) and $m=80$, with block entries (panel (c)) and with random entries (panel (d)). Blue edges mean negative weights and red ones represent positive weights, while the edges are clockwise-oriented. In each graph the nodes represent the $n$ variables of the VAR model, and a clockwise-oriented edge from node $j$ to node $i$ represents a non-null coefficient for the variable $y_{j,t-1}$ in the $i$-th equation of the VAR. Blue edges represent negative coefficients, while red edges represent positive coefficients.}
  \label{graphXi}
 \end{figure}

As one can see from Fig. \ref{graphXi}, two experimental settings have been considered for $B$, which reflect two typical network configurations detected in many empirical studies. The first setting represents a system which is robust to random failures, but vulnerable to targeted attacks.  We consider many nodes (financial institutions) with low degree, and a small number of nodes, known as hubs \citep[e.g.][]{Barabasi1999,Acemoglu2012}, or systemically important institutions \citep[e.g.][]{BilGetLoPel12,Battiston2012}, with a number of linkages that exceeds the average. The presence of hubs characterizes periods of high systemic risk and  when the default of the largest institution (hubs) can lead to the breakdown of the whole system. Hubs have been detected in the S\&P100 return networks \citep[e.g.,][]{BiaBilCasGui15} and in EuroStoxx-600 realized volatility networks \citep[e.g.,][]{BilAheCa09}. Also, we assume nodes cluster into groups internally densely connected with no connections among them. This feature is known as community structure, or modularity \citep{newman2006,newman2008} and has been detected not only in financial-return networks, but also in bank-firm \citep{bargigli2013}, regional banks \citep{puliga2016} and inter-bank market \citep{desouza2016}, networks. The presence of communities structure is a relevant features, since it can prevent the spread of contagion to the whole system. The setting described corresponds to a block-diagonal matrix $B$ with $(4\times 4)$ blocks $B_j$ ($j=1,\ldots,m/4$) on the main diagonal representing the communities, and i.i.d. elements within each block, $b_{lk,j} \sim \mathcal{U}(-1.4,1.4)$, $l,k=1,\ldots,4$. The matrix $B$ is checked for the weak stationarity condition of the VAR.\\
\indent The second setting represents a stable financial or economic condition, which reflects in a network with few connections, randomly distributed across node pairs. This network configuration has been observed during periods of low systemic risk. More specifically, we consider a random matrix $B$ of dimension $(80\times 80)$, select randomly $150$ elements and draw their values from the uniform $\mathcal{U}(-1.4,1.4)$. The remaining $6250$ elements are set to zero. The matrix generated is then checked for weak stationarity. In addition to the two settings previously described, we consider a VAR(4) parametrized on the posterior mean of the macroeconomic application presented in Section \ref{EmpApp}. The network configuration at each lag, given in Fig. \ref{graph_OECD_Lag}, exhibits edge heterogeneity and hubs.

The burn-in period and the thinning rate have been chosen following a graphical dissection of the posterior progressive averages and the standard convergence MCMC diagnostics. Table \ref{Coda_lambda} shows the diagnostics averaged over 50 independent experiments.

\begin{table}[t]
\centering
\begin{tabular}{lcccccc}
\hline
& CD & KS & \multicolumn{2}{c}{INEFF} & \multicolumn{2}{c}{ACF(10)} \\
& & & before & after & before & after \\
\hline
$m = 20$ & -0.629 & 0.3667 & 13.5092 & 7.0346 & 0.1724 & 0.0188\\
$m = 40$ & -1.197 & 0.002 & 21.3259 & 8.5001  & 0.3366 & 0.1334  \\
$m = 80$ block & -1.354 & 0.5806 & 20.1303 & 9.6897 & 0.3147 &  0.1229 \\
$m = 80$ random & 2.451 & 0.0158 & 17.6933 & 5.8095 & 0.267 & 0.0185 \\
\hline
\end{tabular}
\caption{Geweke (CD) and Kolmogorov-Smirnov (KS) convergence test; inefficiency factor (INEFF) and autocorrelation (ACF) computed at lag $10$ for the $\mathcal{L}^2$ norm of $\lambda_{ij}$ for each experiments on the whole (before) and thinned (after) MCMC samples. All quantities are averages over 50 independent experiments.}
\label{Coda_lambda}
\end{table} 
Since in a data augmentation framework, the mixing of the MCMC may depend on the autocorrelation in both parameter and latent variable samples, we focus on $\lambda_{ij}$, $u_{ij}$ and $\beta_{ij}$. The diagnostic for $\lambda_{ij}$, after removing $500$ burn-in iterations, indicates the chain has converged. The inefficiency factor and the autocorrelation at lag $10$ on the original sample suggest high levels of dependence in the sample, which can be mitigated by thinning the chains. We keep only every $5$th sample and obtain substantial reduction of sample autocorrelation. See Supplementary Material for further details.

\begin{figure}[h]
\setlength{\tabcolsep}{4pt}
\begin{tabular}{ccc}
   \includegraphics[width=4.7cm]{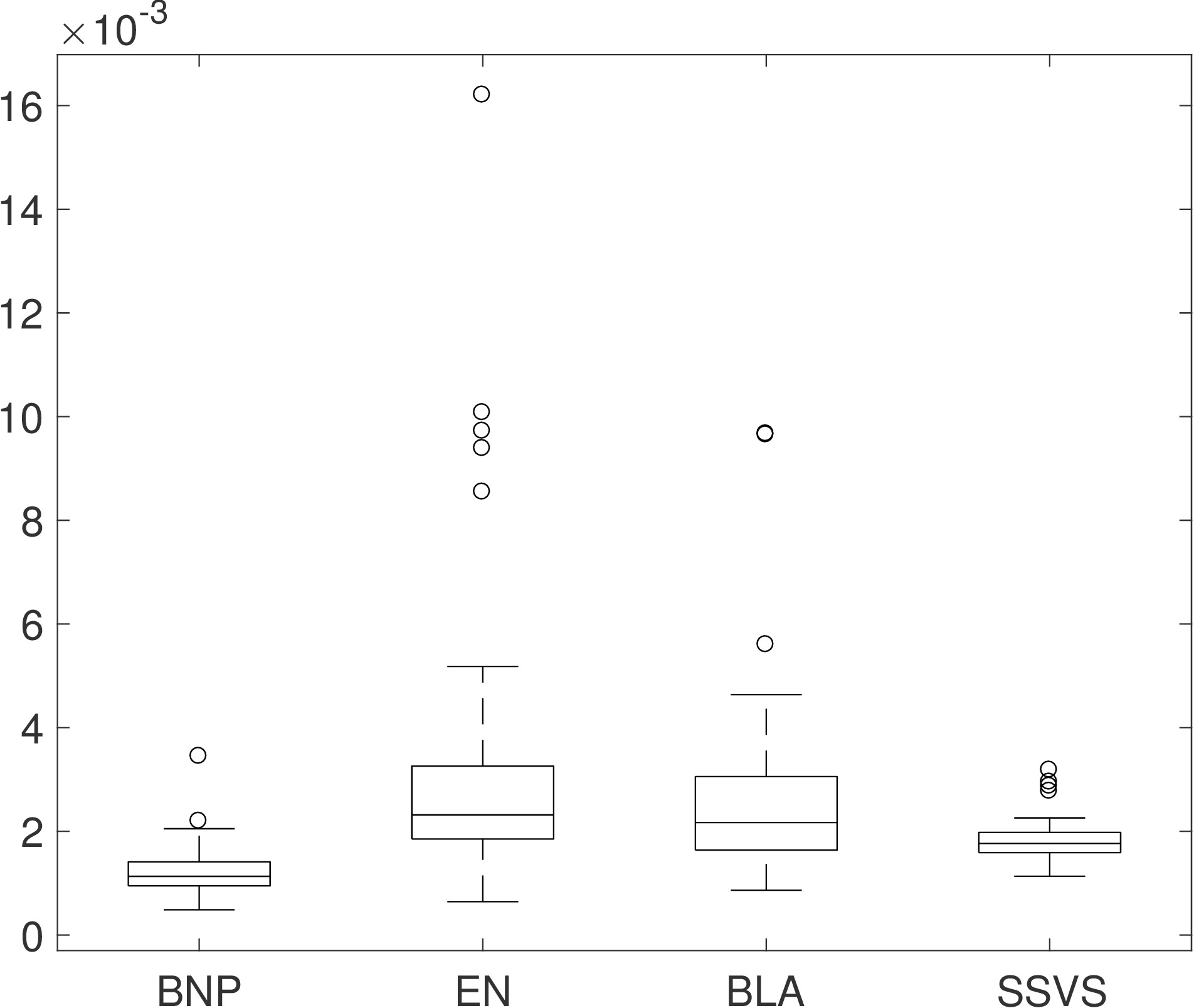} &\includegraphics[width=4.7cm]{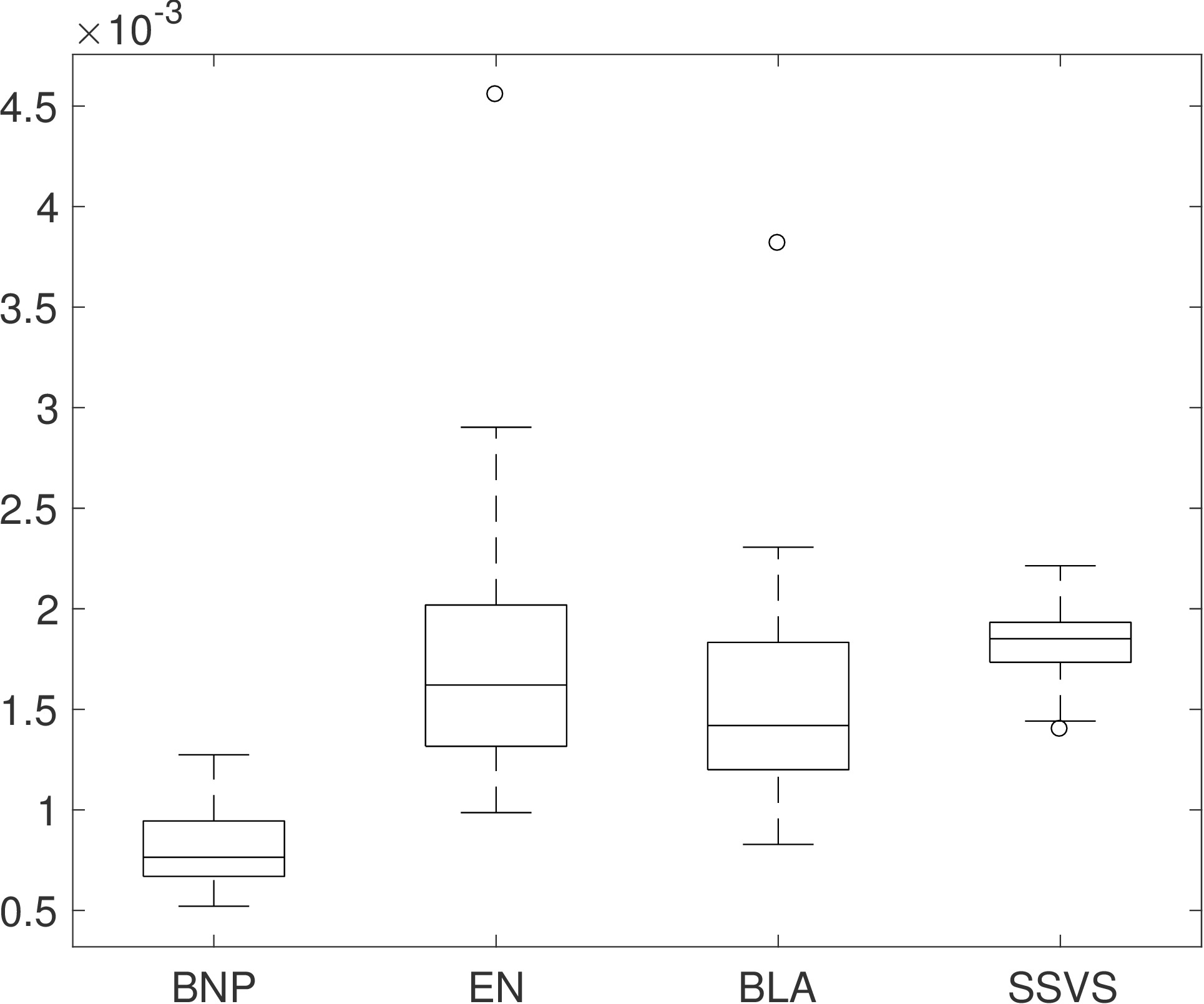} &   \includegraphics[width=4.7cm]{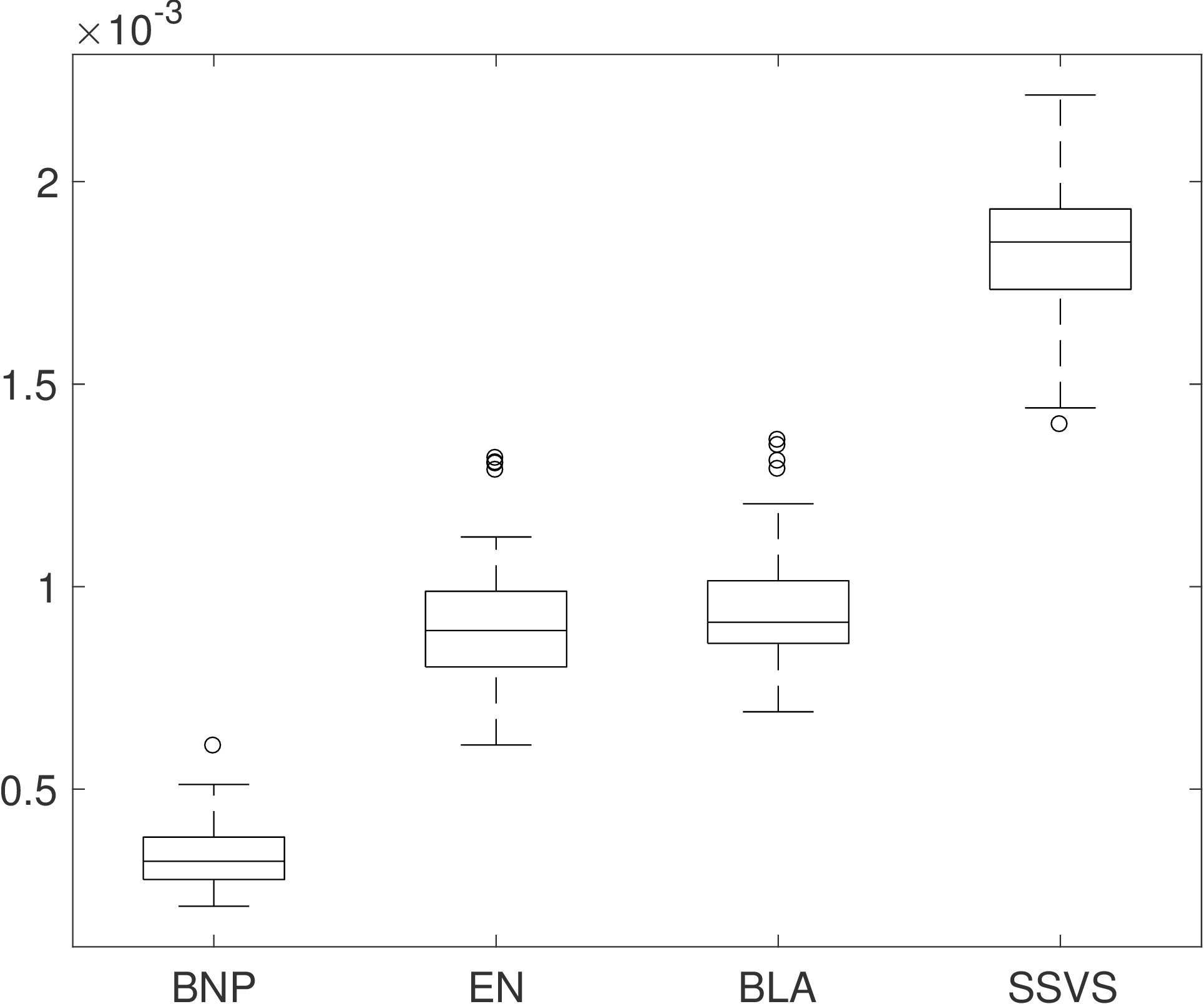} \\
   {\small $m=20$} & {\small $m=40$} &{\small $m=80$ random} \\
\end{tabular}
\caption{Boxplot of Mean Square Deviation (MSD) of the estimated VAR coefficients from their true values based on 50 independent Monte Carlo exercises for our Bayesian nonparametric model (BNP), Elastic-net (EN), Bayesian Lasso (BLA) and Stochastic Search Variable Selection (SSVS). On each box, the central mark indicates the median, while the bottom and top edges of the box are the $25$th and $75$th percentiles, respectively. The whiskers extend to the most extreme points and the outliers are plotted using $\circ$.}
\label{MSD_Box}
\end{figure}

We compare our prior (BNP) with the Bayesian Lasso (B-Lasso, \cite{ParCas05}), the Elastic-net (EN, \cite{ZouHa05}) and Stochastic Search Variable Selection (SSVS) of \cite{GeoSunN08}. For the SSVS, we assume the default hyperparameters values $\tau_1^2=0.0001$, $\tau_2^2=4$ and $\pi=0.5$. Following \cite{Koro16}, we use the Mean Square Deviation (MSD) for measuring the performance of the four different priors. See Supplementary Material for a more detailed description. For each parameter setting we generate $50$ independent datasets and estimate the models under comparison.  Fig. \ref{MSD_Box} shows the quartiles and median of the MSD statistics based on $50$ Monte Carlo experiments by means of box plots. In all settings, it is likely that the BNP-Lasso works better compared to other shrinkage methods in recovering the true VAR coefficients since the interquartile ranges (i.e. the boxes) do not overlap. However, one can better appreciate a superior relative performance of the BNP-Lasso when model dimension increases (from $m=20$ (left) to $m=80$ (right)). 
In our simulation results, the BNP-Lasso shrinking effect is a midway between SSVS and Bayesian Lasso. It is smaller with respect to B-Lasso and Elastic-Net and comparable to SSVS when the true $\beta_{ij}$ is nonnull and it is larger than SSVS when $\beta_{ij}=0$. The reduction in MSD is largely a result of decreased bias. The BNP-Lasso clustering effect tends to identify the correct coefficient clustering reducing the bias. In the Supplementary Material, the comparison in other settings (Fig. S.8-S.11) confirms our findings. Moreover, Fig. S.12 shows that all shrinkage methods are over-performing the Minnesota prior, thus confirming the results given in \cite{KorobilisPettenuzzo2018} for a VAR of dimension $m=20$. Our findings extend their results to higher dimensions ($m=40$ and $m=80$) and to other shrinkage methods (BNP-Lasso, Elastic-Net). Also, we find that the performance of the Minnesota prior deteriorates when the model dimension increases.


\section{Empirical Applications}
\label{EmpApp}
\subsection{Measuring business cycle connectedness} 
Following the literature on international business cycles we consider a multi-country macroeconomic dataset \citep[e.g., see][]{Kose03,Francis12,KauSch12} and extract a network of linkages between the cycles of the OECD countries, by applying BNP-Lasso VAR.

We consider the quarterly GDP growth rate (logarithmic first differences) from 1961:Q1 to 2015:Q2, for a total of $T=215$ observations. Due to missing values in some series, we choose a subset of high industrialised  countries: Rest of the World (Australia, Canada, Japan, Mexico, South Africa, Turkey, United States) and 
Europe (Austria, Belgium, Denmark, Finland, France, Germany, Greece, Ireland, Iceland, Italy, Luxembourg, Netherlands, Norway, Portugal, Spain, Sweden, Switzerland, United Kingdom).

\begin{figure}[t]
\centering
\begin{tabular}{cc}
 \includegraphics[height=4cm, width=5.9cm]{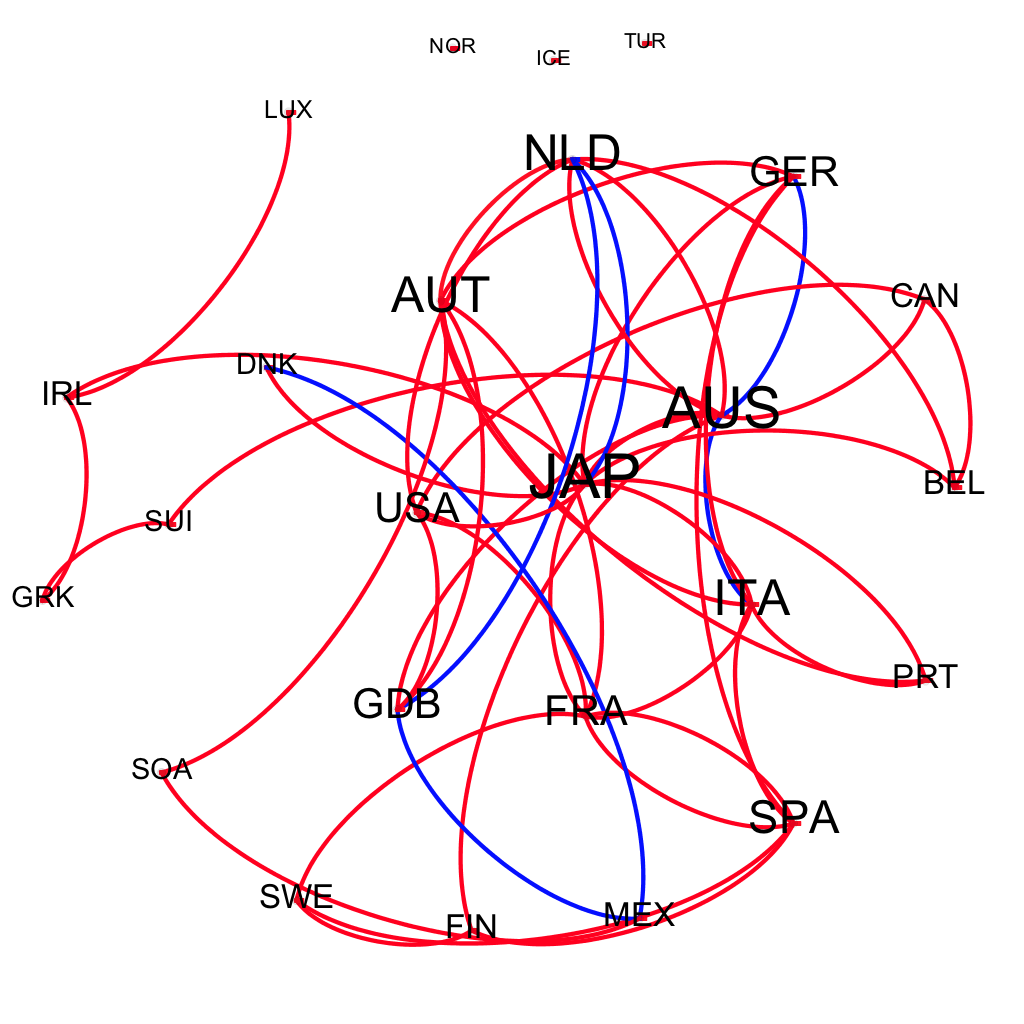} &
 \includegraphics[height=4cm, width=5.9cm]{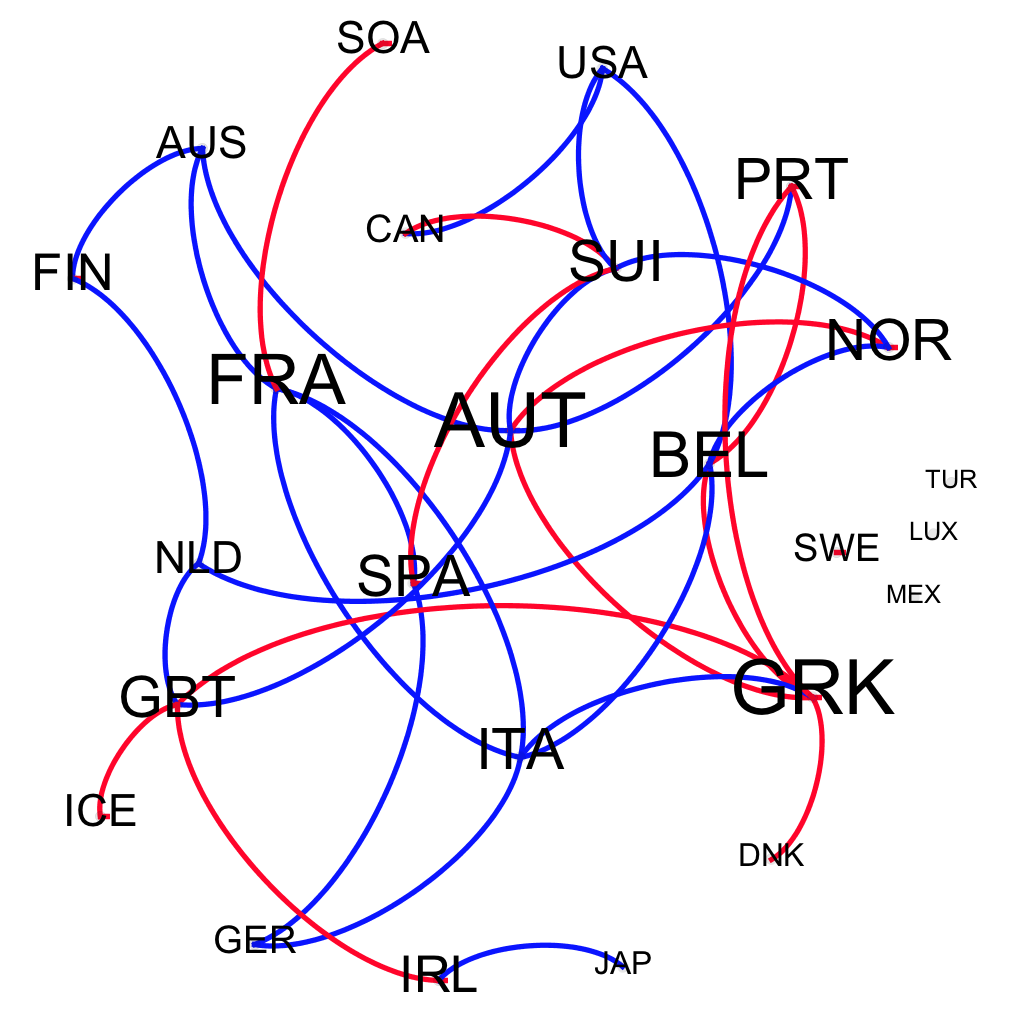} \\
 {\small (a) lag $t-1$ }& {\small (b)  lag $t-2$} \\
\includegraphics[height=4cm, width=5.9cm]{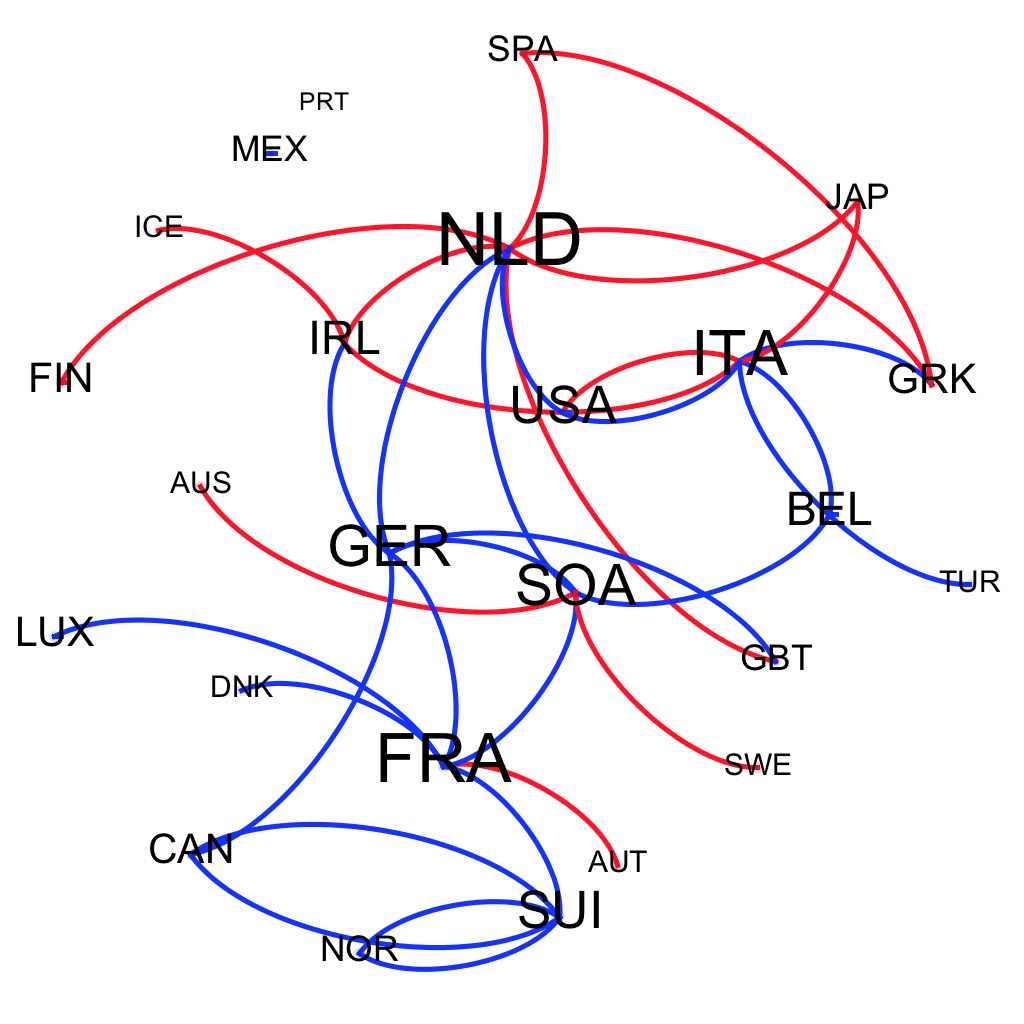}&
\includegraphics[height=4cm, width=5.9cm]{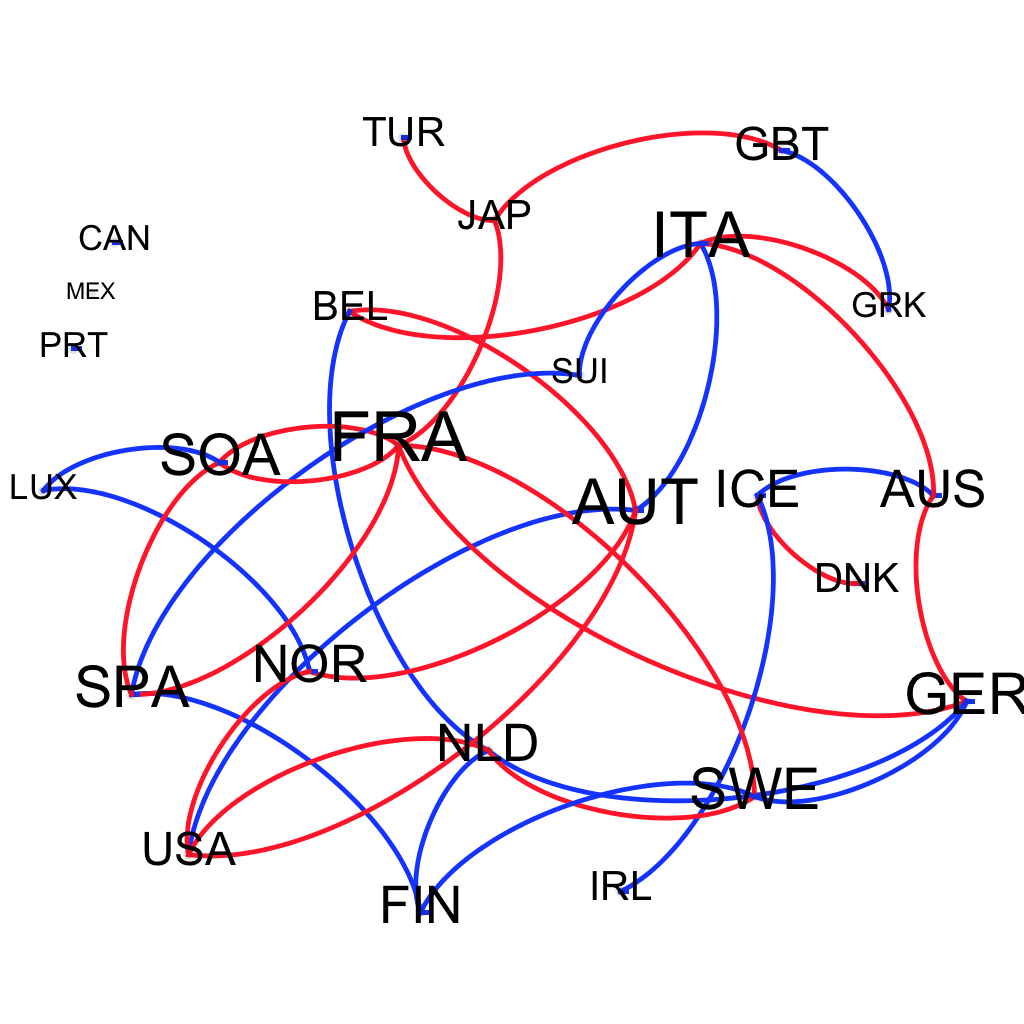} \\
{\small (c) lag $t-3$} & {\small (d) lag $t-4$} \\
\end{tabular}
 \caption{Weighted Granger networks for the GDP growth rates of 25 OECD countries for the sample period 1961Q1 to 2015Q2. We consider different lags: (a) $t-1$, (b) $t-2$, (c) $t-3$, (d) $t-4$. Node size indicates node degree, blue edges represent negative weights and red ones positive weights. At the lag $l$, clockwise-oriented edge from node $j$ to node $i$ represents a non-null coefficient for the variable $y_{j,t-l}$ in the $i$-th equation of the VAR.}
 \label{graph_OECD_Lag}
 \end{figure}

The posterior probability of a macroeconomic linkage is $0.13$, which provides evidence of sparsity in the network and indicates that a small proportion of linkages is responsible for connectedness in business cycle risk. The posterior number of clusters and the co-clustering matrix (see Supplementary Material) reveals the existence of three levels of linkage intensity, customarily called ``negative'', ``positive'' and ``strong positive''. Fig. \ref{graph_OECD_Lag} draws the coloured networks at different time lags with negative (blue) and positive (red) weights.  In a multilayer graph, each intensity level (or edge colour) identifies the set of nodes and edges belonging to a specific layer. The network connectivity and nodes centrality can be investigated either at the global level, or for each single layer.

The network topology analysis in Table \ref{TGraph} reveals a significant connectivity (graph density and average degree) and complexity (average path length) at all lags. Consequently, each lag-specific network contribute to the build up of the global connectedness as shown in Eq. \eqref{Decom}. It becomes important to study the dynamical structure of the directional connectedness received from other countries (in-degree) or transmitted to other countries (out-degree). The multilayer analysis (coloured edges in Fig. \ref{graph_OECD_Lag} and Table \ref{TGraph}) will be used to identify the countries, who mostly contribute at each lag to the system-wide connectedness \citep{DiebYilm14}.\\
\indent At the first two lags, core European countries (Austria, Belgium, Finland, France, Germany, Netherlands) are mainly receivers, whereas the periphery European countries (Greece, Ireland, Italy, Portugal, Spain) transmit the highest percentage of shocks. In the other lags, core countries receive and transmit a high percentage of shocks.\\
\indent Decomposing the degree by the intensity levels (see Eq. \eqref{dec}), we see that the average degree at the first lag ($2.92$) is mainly driven by positive and strong positive edge intensities ($2.64$). The $88\%$ of the linkages provides a positive contribution to the connectedness and originates from Japan (the central country) and European countries. The most central country in the blue layer is the Netherlands, those negative linkages could contribute to reduce the total connectedness. \\
\indent At the second lag, the red and blue layers have similar average degree. Austria is central in the overall network exhibiting both negative and positive linkages. Nevertheless, if one considers only positive linkages, the central country is Greece.\\
\indent At the third and fourth lag, negative linkages prevail over positives. European countries have positive linkage strength within them, thus contributing to a marginal increase of the total connectedness. A reduction of connectedness comes from the negative linkages within rest-of-the-world countries and between them and European countries.

\begin{table}[t]
\centering
\begin{small}
\def\arraystretch{0.8}
\begin{tabular}{lcccc}
\hline
 &  Links & Avg Degree & Density & Avg Path Length  \\
\hline
$t-1$ & 73 & 2.92 & 0.122 & 3.423 \\
$t-1$ blue & 7 & 0.28 & 0.012 & 1.143 \\
$t-1$ red & 66 & 2.64 & 0.11 & 2.587 \\
\hline
$t-2$ & 45 & 1.80 & 0.075 & 3.211 \\
$t-2$ blue & 22 & 0.88 & 0.037 & 2.634 \\
$t-2$ red & 23 & 0.92 & 0.038 & 2.033 \\
\hline
$t-3$ & 41 & 1.64 & 0.069 & 2.479 \\
$t-3$ blue & 25 & 1.00 & 0.042 & 2.268 \\
$t-3$ red & 16 & 0.64 & 0.027 & 1.667 \\
\hline
$t-4$ & 52 & 2.08 & 0.086 & 2.718 \\
$t-4$ blue & 32 & 1.28 & 0.053 & 1.435 \\
$t-4$ red & 20 & 0.80 & 0.033 & 1.791 \\
\hline
\end{tabular}
\end{small}
\caption{Statistics for single layers networks of the GDP growth networks of 25 OECD countries for the sample period 1961Q1 to 2015Q2. The sub-graphs (layers) are induced by positive (red) and negative (blue) edge intensity. The density The average path length represents the average graph-distance between all pairs of nodes. Connected nodes have graph distance 1.}
\label{TGraph}
\end{table}

In order to validate our model, we study its forecasting abilities. In the comparison in Supplementary Material, we find that BNP-Lasso on this specific dataset over-performs Elastic-Net (EN), Bayesian Lasso (B-Lasso) and SSVS. These findings suggest that the BNP-Lasso is  useful not only for a better understanding of the connectedness but could also be employed for forecasting purposes in macroeconomics \citep[e.g.][]{McCracken2016}. Since forecasting is behind the scope of the paper,  we leave it for further research.

\subsection{Risk Connectedness in European Financial Markets}

\begin{figure}[t]
\centering
\setlength{\tabcolsep}{5pt}
\def\arraystretch{0.3}
\begin{tabular}{cc}
 \includegraphics[height=5cm, width=6.2cm]{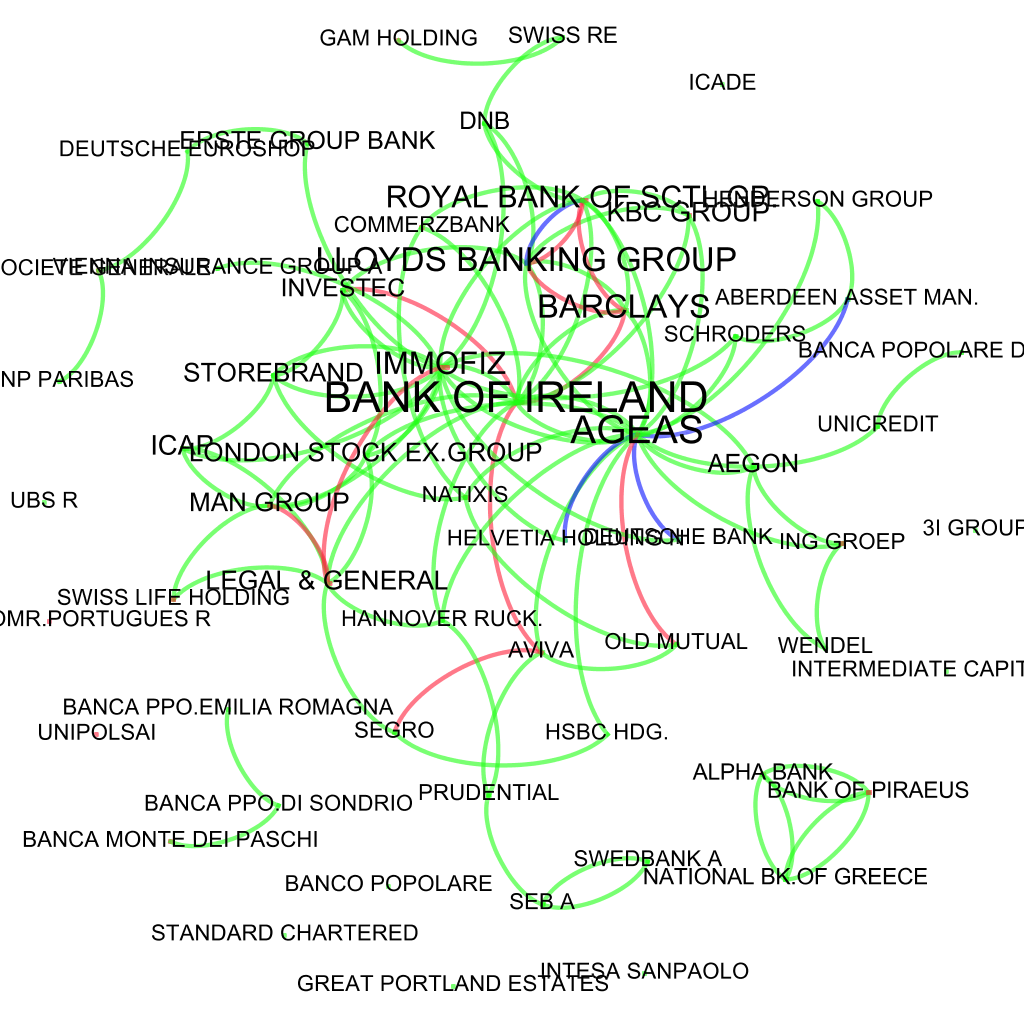} &
    \includegraphics[height=5cm, width=6.2cm]{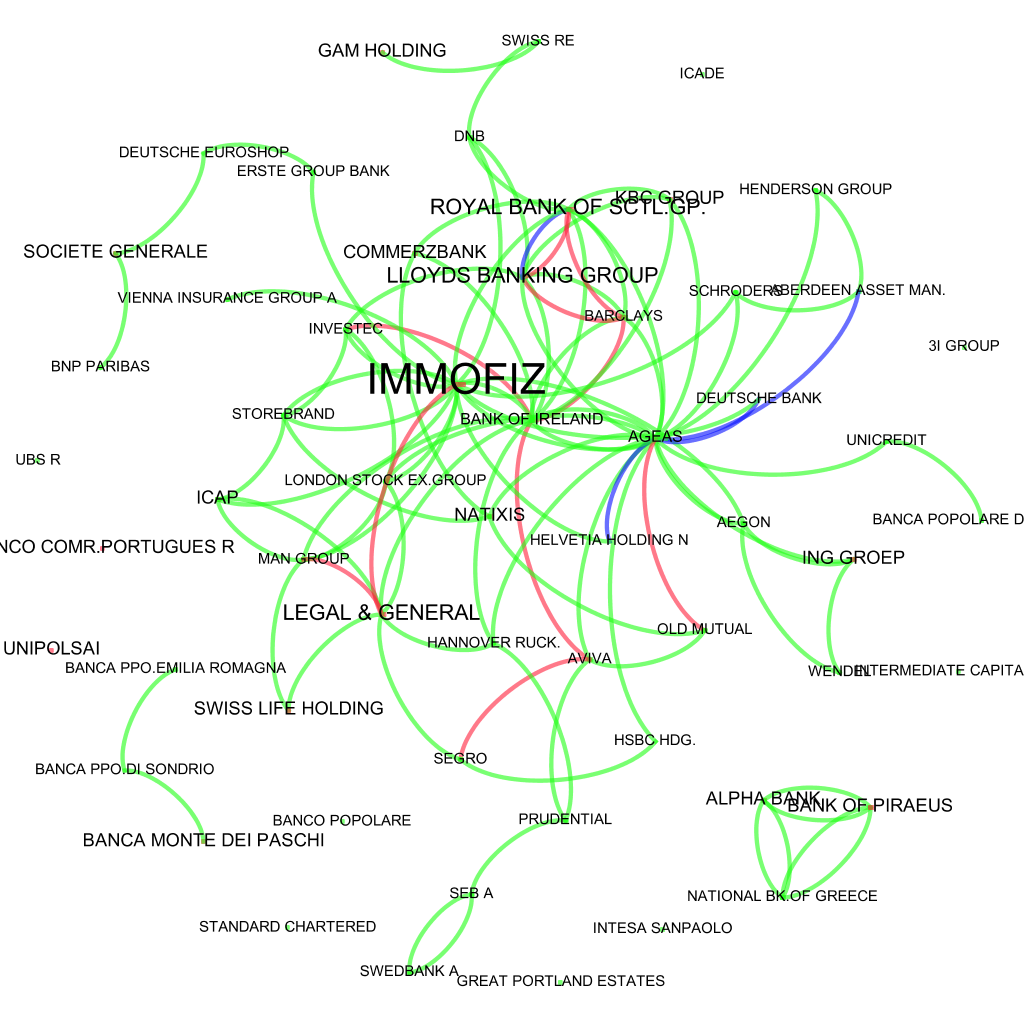} \\
\multicolumn{2}{c}{
  \vspace{-15pt}\includegraphics[height=5cm, width=6.2cm]{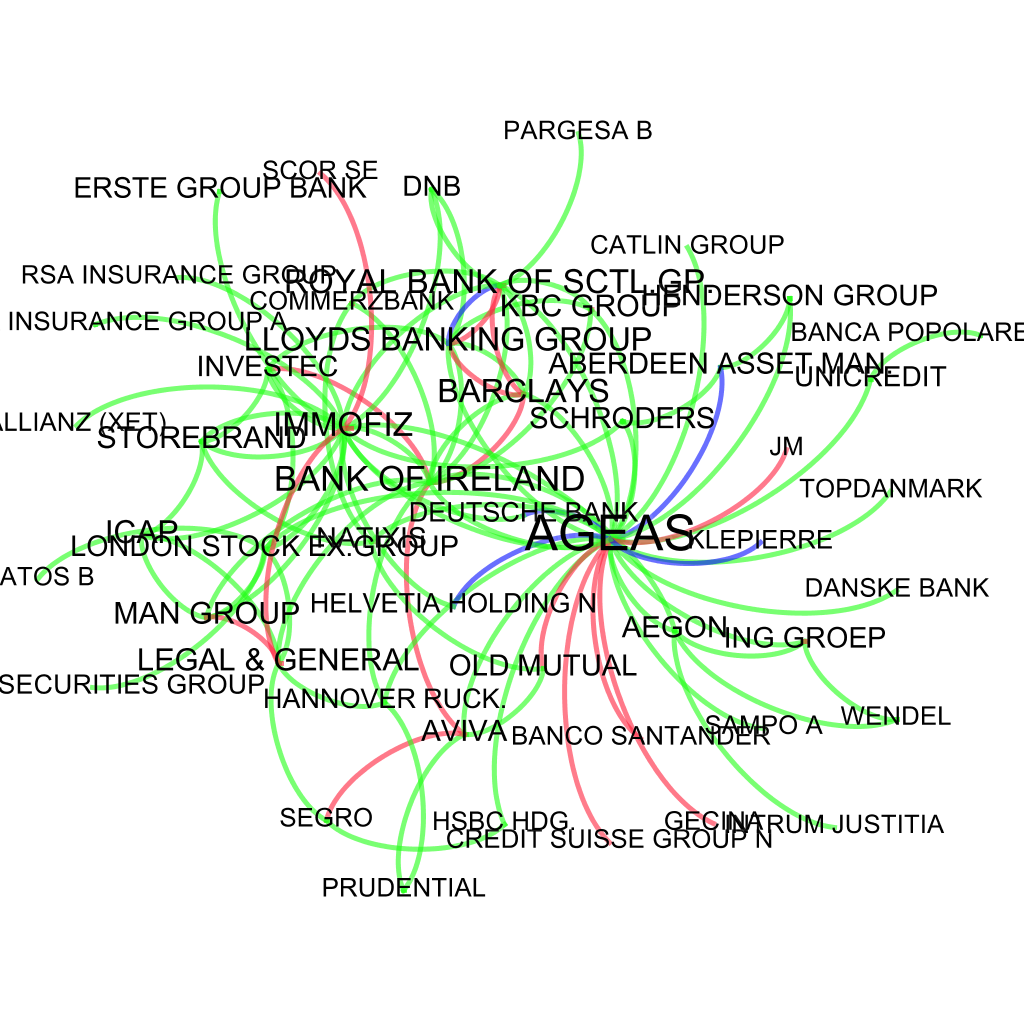}}
\end{tabular}
 \caption{Weighted financial networks of realized volatility for 118 financial institutions (nodes) of the EuroStoxx $600$ from January 3, 2005 to September 19, 2014. We show the pairwise directed edges extracted with BNP-Lasso VAR(1). Edge colours indicate five levels of linkage strengths: ''negative'' (blue), ''weak positive'' (green), ''positive'' (orange) and ''strong positive'' (red). The edges are clockwise-oriented and the node size is based on the node degree centrality (top-left) or eigenvector centrality (top-right). Bottom: AGEAS two-step ego network.}
 \label{graph_RV}
 \end{figure}

Based on the literature on risk connectedness among financial institutions and markets (see \cite{Hautsch15} and \cite{DiebYilm14}), we construct daily realized volatilities using intraday high-low-close price indexes of $118$ institutions of the Euro Stoxx $600$ obtained from Datastream,  from January 3, 2005 to September 19, 2014. The dataset consists of $42$ banks, $31$ financial services, $31$ insurance companies and $22$ real estates from Austria, Belgium, Finland, France, Germany, Greece, Ireland, Italy, Luxembourg, Netherlands, Portugal and Spain. As in \cite{BilAheCa09}, we build the realized volatility as $\text{RV}_t = 0.5\left(\log H_t - \log L_t\right)^2 - \left(2\log 2 -1\right)\left(\log C_t - \log C_{t-1}\right)^2$, where $H_t$, $L_t$ and $C_t$ denote the high, low and closing price of a given stock on day $t$, respectively and consider a VAR(1) model. 

There is a strong evidence of sparsity with a probability of $0.16$ of edge existence (Fig. \ref{graph_RV}). Based on the posterior number of clusters, we identify five levels of linkage strengths, customarily called ''negative'' (blue), ''weak positive'' (green), ''positive'' (orange) and ''strong positive'' (red). Negative linkages imply some institutions react with a volatility decrease to positive volatility shocks, thus reducing systemic risk. Top-left plot in Fig. \ref{graph_RV} shows the coloured realized-volatility network. The node size increases with the node eigenvector centrality, which is evaluated on the overall network without accounting for the linkage intensity. We find that insurance companies and banks (e.g., AGEAS and Bank of Ireland) are central and present different type of linkages (e.g., see AGEAS two-step ego network in the bottom plot). Also, a community structure can be detected with some small communities of same countries banks, such as the one related to the Italy (Monte dei Paschi and other Popular Banks) and Greece (Alpha Bank, National Bank of Greece, Bank of Piraeus). 

Evaluating the node eigenvector centrality in the subgraphs of positive and strong positive linkages, we found that also real estates sector is central (e.g. Immofiz, top-right plot). Our approach thus helps in better understanding the role of financial institutions in the build-up of the systemic risk, revealing different aspects of their activity and in particular for insurance and real estate industries.

\section{Conclusions} \label{Concl}
This paper introduces a novel Bayesian nonparametric Lasso prior (BNP-Lasso) for VAR models, which combines Dirichlet process and Bayesian Lasso priors. The two-stage hierarchical prior allows for clustering the VAR coefficients and for shrinking them either toward zero or random locations, thus inducing sparsity in the VAR coefficients. The simulation studies illustrate the good performance of our model in high dimension settings, compared to some existing priors, such as the Stochastic Search Variable Selection, Elastic-net and simple Bayesian Lasso.\\
\indent The BNP-Lasso VAR is well suited for extracting networks which account for some stylized facts observed in financial and macroeconomic networks such as community structures and linkage heterogeneity. The linkages partition can be easily obtained with the BNP-Lasso framework and the extraction of a coloured network allows for analyzing the centrality of the nodes in the intensity-specific subgraphs.\\
\indent In the macroeconomic (financial) application, our BNP-Lasso VAR helps in understanding the centrality of countries (institutions) and their role in the build-up of risk. In all applications, there is evidence of network sparsity and of different levels of linkage intensity, meaning that few linkages could be responsible for the connectedness. We show evidence that some nodes are central when the intensity level is considered, even if they are not central in the network. Also, few nodes possess linkages with different intensities whereas most of the nodes have only one type of linkages.



%



\bibliographystyle{apalike}
\bibliography{BiblioRevised}

\clearpage

\renewcommand{\thesection}{A}
\renewcommand{\theequation}{A.\arabic{equation}}
\renewcommand{\thefigure}{A.\arabic{figure}}
\renewcommand{\thetable}{A.\arabic{table}}
\setcounter{table}{0}
\setcounter{figure}{0}
\setcounter{equation}{0}

\section{Further details on prior specification}
\label{AppPrior}
\subsection{Normal-Gamma distribution}
A normal-gamma random variable $X \sim \mathcal{N}\mathcal{G}(\mu,\gamma,\tau)$ has probability density function
\begin{align}
f(x|\mu,\gamma,\tau)=\frac{\tau^{\frac{2\gamma+1}{4}}|x-\mu|^{\gamma-\frac{1}{2}}}{2^{\gamma-\frac{1}{2}}\sqrt{\pi}\Gamma(\gamma)}K_{\gamma-\frac{1}{2}}(\sqrt{\tau}|x-\mu|), \notag
\end{align}
where $K_{\gamma}(\cdot)$ represents the modified Bessel function of the second kind with index $\gamma$, $\mu \in \mathbb{R}$ is the location parameter, $\gamma>0$ is the shape parameter and $\tau>0$ is the scale parameter.
The normal-gamma distribution has the double exponential distribution as a special case for $\gamma=1$ and can be represented as a scale mixture of normals, 

\begin{equation}
\mathcal{N}\mathcal{G}(x|\mu,\gamma,\tau)= \int_0^{+\infty} \mathcal{N}(x| \mu,\lambda)  \Ga(\lambda|\gamma, \tau/2) d\lambda, 
\label{3App}
\end{equation}
where $\Ga(\cdot|a,b)$ denotes a gamma distribution with mean $a/b$ and variance $a/b^2$.

\subsection{Gamma scale-shape distribution}
 
\begin{figure}[h!]
\centering
{\includegraphics[width=7cm]{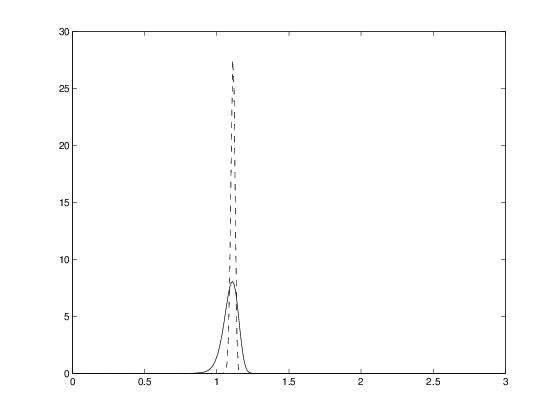}}
\caption{Probability density function, $g(x)$, for sparse (dashed line) and non-sparse (solid line) case, respectively.}
\label{Fig1}
\end{figure}

A Gamma scale-shape random vector $(X,Y) \sim \mathcal{G}S(\nu,p,s,n)$ has pdf
\begin{align}
g(x,y|\nu,p,s,n) \propto \, \tau^{\nu x-1}p^{x-1}\exp\{-s y\}\frac{1}{\Gamma(x)^{n}}, \label{gzero}
\end{align}
with parameters $\nu>0$, $p>0$, $s>0$ and $n>0$ (see \cite{Miller1980}). The pdf in equation \eqref{gzero} factorizes as $g(x,y)=g(y|x)g(x)$, where
\begin{equation*}
g(y|x)=\frac{g(x,y)}{g(x)}=\frac{ \tau^{\nu x-1}e^{-sy}}{\Gamma(\nu x)}s^{\nu x}
\end{equation*}
that is a density of a $\mathcal{G}a(\nu x,s)$, and 
\begin{equation*}
g(x)=\int_0^\infty g(x,y) dy =C \frac{\Gamma(\nu x)}{\Gamma(x)^{n}}\frac{p^{x-1}}{s^{\nu x}}
\end{equation*} 
is the marginal density with normalizing constant $C$ such that $\int_0^\infty g(x) dx = 1$. We show in Figure \ref{Fig1} the density function, $g(x)$, for the two parameter settings used in the empirical application: $v=30$, $s=1/30$, $p=0.5$ and $n=18$ (dashed line) and $v=3$, $s=1/3$, $p=0.5$ and $n=10$ (solid line).

\subsection{Hyper-inverse Wishart Distribution}

An Hyper-inverse Wishart random variable $\Sigma \sim \mathcal{H}IW_G(b,L)$ has pdf 
\begin{equation}
p(\Sigma)=\prod_{P\in \mathcal{P}} p(\Sigma_{P}) \left(\prod_{S\in \mathcal{S}} p(\Sigma_{S})\right)^{-1}, \notag
\end{equation} 
where $\mathcal{S}=\{S_1,\dots, S_{n_S}\}$ and $\mathcal{P}=\{P_1,\dots,P_{n_P}\}$ are the set of separators and of prime components, respectively, of the graph $G$. In particular, $b$ is the degree of freedom, $L$ is the scale matrix and 
\begin{equation}
p(\Sigma_{P})\propto \, |\Sigma_{P}|^{-(b+2\text{Card}(P))/2} \exp{\left\{-\frac{1}{2}\hbox{tr}(\Sigma_{P}^{-1} L_P)\right\}}, \notag
\end{equation} 
with $L_{P}$ the positive-definite symmetric diagonal block of L corresponding to $\Sigma_{P}$.

\subsection{Dirichlet Process prior}
The Dirichlet Process, $\text{DP}(\tilde{\alpha},H)$, can be defined by using the stick-breaking representation (\cite{Sethuraman}) given by:
\begin{equation}
\mathbb{P}_i (\cdot)= \sum_{j=1}^{\infty} w_{ij} \delta_{\{\bm{\theta}_{ij}\}}(\cdot), \, \quad i=1,\dots,M.
\notag
\end{equation} 
Following the definition of the dependent stick-breaking processes (see \cite{MacEachern1999} and \cite{MacEachern2001}), the atoms $\theta_{ij}$ are i.i.d. sequences of random elements with probability distribution $H$ ($\theta_{ij} \overset{\text{i.i.d.}}{\sim} H$); and the weights $w_{ij}$ are determined through the stick-breaking construction, for $j=1$, $w_{i1}=v_{i1}$ and for $j>1$
\begin{equation}
w_{ij}=v_{ij}\prod_{k=1}^{j-1} (1-v_{ik}), \quad i=1,\dots,M
\notag
\end{equation}
with $v_j=(v_{1j},\dots,v_{Mj})$ independent random variables taking values in $[0,1]^M$ distributed as a $\Beta(1,\tilde{\alpha})$ such that $\sum_{j\ge 1} w_{ij}=1$ almost surely for every $i$.


\renewcommand{\thesection}{B}
\renewcommand{\theequation}{B.\arabic{equation}}
\renewcommand{\thefigure}{B.\arabic{figure}}
\renewcommand{\thetable}{B.\arabic{table}}
\setcounter{table}{0}
\setcounter{figure}{0}
\setcounter{equation}{0}

\section{Proof of the results in the paper}
\label{App_Res}

\begin{proof}[Proof of result in Eq.  \eqref{InfMix}]

Let $K(\bm{\beta}_i|\bm{\theta}_i)$ be the joint density kernel of the sequence $\beta_{ij}\overset{ind}{\sim}\mathcal{N}\mathcal{G}(\beta_{ij}|	\bm{\theta}_{ij}^{*})$, $j=1,\ldots,n_i$ and $\bm{\theta}_{ij}^{*}|\mathbb{Q}_i\overset{i.i.d.}{\sim}\mathbb{Q}_i,$ the atoms of the random probability measure $\mathbb{Q}_i$. Let $f_i(\bm{\beta}_i|\mathbb{Q}_i)$ be the random probability density function obtained by integrating out the random measure $\mathbb{Q}_i$ (\cite{Lo1984}):
\begin{equation}
f_i(\bm{\beta}_i|\mathbb{Q}_i)= \int K(\bm{\beta}_i| \bm{\theta}_i) \mathbb{Q}_i(d\bm{\theta}_i).
\label{7}
\end{equation}

\noindent Using the definition of $\mathbb{Q}_i$ as a convex combination of a sparse component and a DPP in equation \eqref{7}, we obtain 
\begin{equation}
f_i(\bm{\beta}_i|\mathbb{Q}_i)= \pi_i \int K(\bm{\beta}_i|\bm{\theta}_i) \mathbb{P}_0(d\bm{\theta}_i)+(1-\pi_i)\int K(\bm{\beta}_i|\bm{\theta}_i) \mathbb{P}_i(d\bm{\theta}_i) \label{Inf_Mix_DPP}
\end{equation}
We use the stick-breaking representation of the DPP (Appendix \ref{AppPrior}) and get
\begin{align}
f_i(\bm{\beta}_i|\mathbb{Q}_i)&=\pi_i\int \mathcal{N}\mathcal{G}(\bm{\beta}_i|\bm{\mu},\bm{\gamma},\bm{\tau}) \mathbb{P}_{0}(d(\bm{\mu},\bm{\gamma},\bm{\tau}))+(1-\pi_i)\int \mathcal{N}\mathcal{G}(\bm{\beta}_i|\bm{\mu},\bm{\gamma},\bm{\tau}) \mathbb{P}_{i}(d(\bm{\mu},\bm{\gamma},\bm{\tau}))\notag \\
&= \pi_i\mathcal{N}\mathcal{G}(\bm{\beta}_i|0,\gamma_0,\tau_0) + (1-\pi_i)\sum_{k=1}^\infty w_{ik} \mathcal{N}\mathcal{G}(\bm{\beta}_i|\mu_{ik},\gamma_{ik},\tau_{ik}) \label{InfMixApp}
\end{align}
By defining, 
\[
\check{w}_{ik}=\left\{
\begin{array}{ll}
\pi_i,&k=0,\\
(1-\pi_i)w_{ik},&k>0,\\
\end{array}
\right.
\quad
\check{\theta}_{ik}=\left\{
\begin{array}{ll}
(0,\gamma_0,\tau_0),&k=0,\\
(\mu_{ik},\gamma_{ik},\tau_{ik}),&k>0.\\
\end{array}
\right.
\]
we have the infinite mixture representation 
\begin{equation*}
f_i(\bm{\beta}_i|\mathbb{Q}_i) = \sum_{k=0}^\infty \check{w}_{ik} \mathcal{N}\mathcal{G}(\bm{\beta}_i|\check{\theta}_{ik}).
\end{equation*}
\end{proof}  

\begin{proof}[Proof of result in Eq. \eqref{JointPos}]
We introduce a set of slice latent variables, $u_{ij}, j=1,\dots,n_i$, which allows us to represent the full conditional of $\beta_{ij}$ as follows,
\begin{align}
f_i(&\beta_{ij},u_{ij}| (\mu_i,\gamma_i,\tau_i),w_i)=\pi_i \sum_{k=0}^{\infty} \mathbb{I}(u_{ij}<\tilde{w}_{ik})\mathcal{N}\mathcal{G}(\beta_{ij}|(0,\gamma_{ik},\tau_{ik})) + \notag \\
&+ (1-\pi_i) \sum_{k=1}^{\infty} \mathbb{I}(u_{ij} <w_{ik})\mathcal{N}\mathcal{G}(\beta_{ij}|\mu_{ik},\gamma_{ik},\tau_{ik}) \notag \\
&= \pi_i \mathbb{I}(u_{ij}<\tilde{w}_{0}) \mathcal{N}\mathcal{G}(\beta_{ij}|(0,\gamma_{0},\tau_{0})) + (1-\pi_i) \sum_{k=1}^{\infty} \mathbb{I}(u_{ij} <w_{ik})\mathcal{N}\mathcal{G}(\beta_{ij}|\mu_{ik},\gamma_{ik},\tau_{ik}), \notag
\end{align}
where $\tilde{w}_{ik}=\tilde{w}_0=1$ if $k=0$ and $\tilde{w}_{ik}=0$ for $k>0$ and, for simplicity of notations, we denote $(0,\gamma_{i0},\tau_{i0})=(0,\gamma_0,\tau_0)$.

\noindent The slice variables allow us to represent the infinite mixture as a finite mixture conditionally on a random number of components. More specifically, define the set
\begin{align}
\mathcal{A}_{w_i}(u_{ij})&=\{k: u_{ij}<w_{ik}\},\qquad j=1,\dots,n_i,\notag 
\end{align}
then it can be proved that the cardinality of $\mathcal{A}_{w_i}$ is almost surely finite. Posterior draws for $u_{ij}$ are easily obtained by slice sampling.

\noindent The conditional joint pdf for $\beta_{ij}$ and $u_{ij}$, $f_i(\beta_{ij},u_{ij}| (\mu_i,\gamma_i,\tau_i),w_i)$, is equal to
\begin{align}
\pi_i\mathbb{I}(u_{ij}<\tilde{w}_{0}) \mathcal{N}\mathcal{G}(\beta_{ij}|0,\gamma_0,\tau_0) +(1-\pi_i) \sum_{k\in \mathcal{A}_{w_i}(u_{ij})}\mathcal{N}\mathcal{G}(\beta_{ij}|\mu_{ik},\gamma_{ik},\tau_{ik}). \notag
\end{align}
We iterate the data augmentation principle and for each $f_i$ we introduce two allocation variables $\xi_{ij}$ and $d_{ij}$, associated with the sparse and non-sparse components, respectively, of the random measure $\mathbb{Q}_i$. The joint pdf is
\begin{align}
f_i(\beta_{ij},u_{ij},d_{ij}, \xi_{ij})& = \left(\pi_i\mathbb{I}(u_{ij}<\tilde{w}_{d_{ij}})\mathcal{N}\mathcal{G}(\beta_{ij}|0,\gamma_0,\tau_0)\right)^{1-\xi_{ij}} \cdot\notag \\
&\quad\left((1-\pi_i)\mathbb{I}(u_{ij}<w_{ld_{ij}}) \mathcal{N}\mathcal{G}(\beta_{ij}|\mu_{id_{ij}},\gamma_{id_{ij}},\tau_{id_{ij}})\right)^{\xi_{ij}}. \notag
\end{align}

\noindent From \eqref{3}, we de-marginalize the Normal-Gamma distribution by introducing a latent variable $\lambda_{ij}$ for each $\beta_{ij}$ and conclude that the joint distribution is
\begin{align}
f_i(&\beta_{ij},\lambda_{ij},u_{ij},d_{ij},\xi_{ij})=\left(\mathbb{I}(u_{ij}<\tilde{w}_{d_{ij}})\mathcal{N}(\beta_{ij}|0,\lambda_{ij})\mathcal{G}a(\lambda_{ij}|\gamma_0,\tau_0/2)\right)^{1-\xi_{ij}}\cdot \notag\\
&\,\, \left(\mathbb{I}(u_{ij}<w_{id_{ij}}) \mathcal{N}(\beta_{ij}|\mu_{id_{ij}},\lambda_{ij})\mathcal{G}a(\lambda_{ij}|\gamma_{id_{ij}},\tau_{id_{ij}}/2)\right)^{\xi_{ij}}\pi_i^{1-\xi_{ij}}(1-\pi_i)^{\xi_{ij}}. \notag
\end{align}
\end{proof}

\begin{proof}[Proof of result in Eq.  \eqref{Decom}]
For $h\le p$, $\Phi_h$ satisfies $ \Phi_h = R_h + B_h$, where $R_h =\sum_{l=1}^{h-1} B_l \Phi_{h-l}$. By the properties of the Hadamard product $\circ$, it follows 
 \begin{align*}
&\left(\mathbf{e}_i'\Phi_h \Sigma \mathbf{e}_j\right)^2 = 
\mathbf{e}_i' \bigl(\left((R_h +B_h)\Sigma \right) \circ \left((R_h +B_h) \Sigma \right) \bigr)  \mathbf{e}_j =\mathbf{e}_i' \left(\left(R_h \Sigma \right) \circ \left( R_h \Sigma \right) \right) \mathbf{e}_j  \\
&+ \mathbf{e}_i'\left(( R_h \Sigma) \circ (B_h \Sigma) \right) \mathbf{e}_j + \mathbf{e}_i'\left( (B_h \Sigma) \circ ( R_h \Sigma) \right)  \mathbf{e}_j + \mathbf{e}_i'\left( (B_h\Sigma) \circ(B_h\Sigma) \right) \mathbf{e}_j  \\
&= \left(\mathbf{e}_i'  \left(R_h \Sigma \right)\mathbf{e}_j \right)^2  + 2 \left(\mathbf{e}_i'R_h \Sigma\mathbf{e}_j\right) \left( \mathbf{e}_i'B_h \Sigma   \mathbf{e}_j \right) + \left( \mathbf{e}_i'B_h\Sigma \mathbf{e}_j \right)^2 
\end{align*}

\end{proof}

\renewcommand{\thesection}{C}
\renewcommand{\theequation}{C.\arabic{equation}}
\renewcommand{\thefigure}{C.\arabic{figure}}
\renewcommand{\thetable}{C.\arabic{table}}
\setcounter{table}{0}
\setcounter{figure}{0}
\setcounter{equation}{0}


\section{Gibbs sampling details}\label{AppAFsCp}
We introduce for $k\ge 1$ the set of indexes of the coefficients allocated to the $k$-th mixture component, $\mathcal{D}_{ik}=\{j\in {1,\dots,n_i} : d_{ij}=k, \xi_{ij}=1 \}$ and the set of the non-empty mixture components, $\mathcal{D}^*=\{k |  \cup_{i} \mathcal{D}_{ik} \ne 0 \}$. The number of stick-breaking components is denoted by $D^*=\max_{i}\{\max_{j\in\{1,\dots,n_i\}} d_{ij}\}$.
As noted by \cite{walker2011}, the sampling of infinitely many elements of $\Theta$ and $V$ is not necessarily, since only the elements in the full conditional distributions of $(D,\Xi)$ are needed and the maximum number is $N^*=\max_i\{N_i^* \}$, where $N_i^*$ is the smallest integer such that $\sum_{k=1}^{N_i^*} w_{ik} > 1-u_i^*$, where $u_i^*=\min_{1\le j\le n_i} \{u_{ij}\}$. 
\subsubsection*{Update the stick-breaking and slice variables $V$ and $U$} 
We treat $V$ as three blocks: $V^*=\{V_k : k \in \mathcal{D}^*\}$, $V^{**}=(v_{kD^*+1},\dots v_{kN^*})$ and $V^{***}=\{V_k: k> N^{*}\}$.
In order to sample $(U,V)$, a further blocking is used:
\begin{itemize}
\item[i)] Sampling from the full conditional of $V^{*}$, by using
\begin{align}
f(v_{ij}|\dots)& \propto \, \mathcal{B}e \left(1+\sum_{j=1}^{n_i} \mathbb{I}(d_{ij}=d,\xi_{ij}=1), \alpha + \sum_{j=1}^{n_i}\mathbb{I}(d_{ij}>d,\xi_{ij}=1)\right). \notag 
\end{align}
for $k\le D^*$. The elements of $(V^{**},V^{***})$ are sampled from the prior $\mathcal{B}e (1, \alpha).$
\item[ii)] Sampling from full conditional posterior distribution of $U$
\begin{equation}
f(u_{ij}|\dots)\propto \left\{
\begin{array}{ll}
\mathbb{I}(u_{ij}<w_{1d_{ij}})^{\xi_{ij}}  & \mathrm{if} \; \xi_{ij}=1, \\
\mathbb{I}(u_{ij}<1)^{1-\xi_{ij}} & \mathrm{if} \; \xi_{ij}=0,
\end{array}
\right.
\notag
\end{equation}

\end{itemize}
\subsubsection*{Update the mixing parameters \texorpdfstring{$\bm{\lambda}$}{lambda}}
Regarding the mixing parameters $\lambda_{ij}$, the full conditional posterior distribution is
\begin{align}
f(&\lambda_{ij}|\dots)\propto \, \lambda_{ij}^{C_{ij}-1} \exp{\left\{-\frac{1}{2}\left[A_{ij}\lambda_{ij} + \frac{B_{ij}}{\lambda_{ij}}\right]\right\}} \propto \, \mathcal{G}i\mathcal{G}(A_{ij},B_{ij},C_{ij}), \notag
\end{align}
where $\mathcal{G}i\mathcal{G}$ denotes a Generalize Inverse Gaussian with $A_{ij}>0$, $B_{ij}>0$ and $C_{ij} \in \mathbb{R}$ 
\begin{align}
A_{ij}= \left[ (1-\xi_{ij})\tau_0 + \xi_{ij} \tau_{id_{ij}}\right],& \qquad B_{ij}=\left[(1-\xi_{ij})\beta_{ij}^2 + \xi_{ij}(\beta_{ij}-\mu_{id_{ij}})^2\right], \notag \\
C_{ij}=\biggl[(1-\xi_{ij}&)\gamma_0 + \gamma_{id_{ij}}\xi_{ij} - \frac{1}{2} \biggr]. \notag
\end{align}
The matrix $\Lambda_i=\hbox{diag}{\{\bm{\lambda}_i\}}$ has the elements of $\bm{\lambda}_i=(\lambda_{i1},\dots,\lambda_{in_i})'$ on the diagonal.

\subsubsection*{Update the atoms \texorpdfstring{$\Theta$}{Theta}}
We consider the sparse and the non-sparse case. In the sparse case, the full conditional distribution of $\mu_0$ is $f(\mu_0|\dots)=\delta_{\{0\}}(\mu_0)$ and the full conditional distribution of $(\gamma_0,\tau_0)$ is:
\begin{small}
\begin{align}
f((\gamma_0,\tau_0)|\dots) \propto \mathcal{G}S\left(\gamma_0,\tau_0|\nu_0+ \sum_{i=1}^M n_{i,0},p_0 \prod_{i=1}^M\prod_{j|\xi_{ij}=0} \lambda_{ij}, s_0+\frac{1}{2} \sum_{i=1}^M \sum_{j|\xi_{ij}=0}\lambda_{ij}, n_0+\sum_{i=1}^M n_{i,0}\right), \notag 
\label{p0}
\end{align}
\end{small}
where we assume $n_{i,0}=\sum_{j=1}^{n_i} (1-\xi_{ij})=n_i-n_{i,1}$ and $n_{i,1}=\sum_{j=1}^{n_i} \xi_{ij}$. In the non-sparse case, we generate samples $(\mu_{ik},\gamma_{ik},\tau_{ik})$, $k=1,\dots,N^*$, by applying a single move Gibbs sampler. The full conditional for $\mu_{ik}$ is
\begin{align}
f&(\mu_{ik}|\dots) \propto \, \mathcal{N}(\mu_{ik}|c,d)\prod_{j|\xi_{ij}=1,d_{ij}=k}\mathcal{N}(\beta_{ij}|\mu_{ik},\lambda_{ij}) \propto \mathcal{N}(\tilde{E}_k,\tilde{V}_{k})  \notag
\end{align} 
with $\tilde{E}_{k}=\tilde{V}_{k}\left(\frac{c}{d} + \sum_{j|\xi_{ij}=1,d_{ij}=k} \frac{\beta_{ij}}{\lambda_{ij}} \right)$ and $\tilde{V}_{k}=\left(\frac{1}{d}+\sum_{j|\xi_{ij}=1,d_{ij}=k}\frac{1}{\lambda_{ij}}\right)^{-1}$, the mean and variance, respectively.
The joint conditional posterior of $(\gamma_{ik},\tau_{ik})$ is:
\begin{small}
\begin{align}
f((\gamma_{ik},\tau_{ik})|\dots)\propto \mathcal{G}S\left(\gamma_{ik},\tau_{ik}|\nu_1+n_{i,1k},p_1\prod_{j|\xi_{ij}=1,d_{ij}=k} \lambda_{ij},s_1+\frac{1}{2}\sum_{j|\xi_{ij}=1,d_{ij}=k}\lambda_{ij}, n_1+n_{i,1k}\right), \notag
\end{align}
\end{small}
for $k\in \mathcal{D}^*$, where $n_{i,1k}=\sum_{j=1}^{n_i} \xi_{ij} \mathbb{I}(d_{ij}=k)$ and from the prior $H$ for $k\notin \mathcal{D}^*$. In order to draw samples from $\mathcal{G}S$ in both cases, we apply a collapsed Gibbs sampler. Samples from $f(\gamma)$ are obtained by a Metropolis-Hastings (MH) algorithm and samples from $f(\tau |\gamma)$ are obtained from a Gamma distribution.

\subsubsection*{Update the coefficients \texorpdfstring{$\bm{\beta}$}{beta}} 
The full conditional posterior distribution of $\bm{\beta}$ is:
\begin{align}
f(\bm{\beta}_i&|\dots)\propto \mathcal{N}_{n_i}(\mathbf{\tilde{v}_i},M_i), \notag 
\end{align}
with mean $\mathbf{\tilde{v}_i}= M_i\left(\sum_{t} X_t'\Sigma^{-1}\mathbf{y}_t + \Lambda_i^{-1} (\bm{\mu}_i^*\odot \bm{\xi}_i)\right)$ and variance $M_i=\left(\sum_t X'_t \Sigma^{-1} X_t + \Lambda_i^{-1}\right)^{-1}$;
 and $\bm{\mu}_i^*=(\mu_{id_{i1}},\dots,\mu_{id_{in_{i}}})'$, $\bm{\xi}_i=(\xi_{i1},\dots,\xi_{in_i})'$.

\subsubsection*{Update the covariance matrix \texorpdfstring{$\Sigma$}{Sigma}}

By using the sets $\mathcal{S}$ and $\mathcal{P}$ as described in Appendix \ref{AppPrior} and a decomposable graph, the likelihood of the graphical Gaussian model can be approximated as the ratio between the likelihood in the prime and in the separator components. Thus, the posterior for $\Sigma$ factorizes as follows:
\begin{align}
p(\Sigma&|\dots) \propto \mathcal{HIW}_G \left(b+T, L + \sum_{t=1}^T (y_t - X'_t\bm{\beta})' (y_t - X'_t\bm{\beta}) \right). \notag
\end{align}
\subsubsection*{Update the graph G}
We apply a MCMC for multivariate graphical models for $G$ (see \cite{JonCar05}) and due to prior independence assumption,
\begin{small}
\begin{align}
p(\mathbf{y}|G)=\iint \prod_{t=1}^T &(2\pi)^{-n/2} |\Sigma|^{-n/2} \exp{\biggl(-\frac{1}{2} (y_t - X_t'\bm{\beta})\Sigma^{-1}(y_t-X_t'\bm{\beta})\biggr)} p(\bm{\beta})p(\Sigma|G) d\bm{\beta} d\Sigma. \notag
\end{align}
\end{small}
Following \cite{JonCar05} we apply a local-move MH based on the conditional posterior $p(G|\dots)$ with an add/delete edge move proposal. 

\subsubsection*{Update the allocation variables $D$ and $\Xi$}  
Sampling from the full conditional of D is obtained from
\begin{align}
P(d_{ij}=d,\xi_{ij}=1|\dots)\propto \frac{(1-\pi_i)\mathcal{N}(\beta_{ij}|\mu_{id},\lambda_{ij}) \mathcal{G}a(\lambda_{ij}|\gamma_{id},\tau_{id}/2)}{\sum_{k\in A_{w_i}(u_{ij})}\mathcal{N}(\beta_{ij}|\mu_{ik},\lambda_{ij}) \mathcal{G}a(\lambda_{ij}|\gamma_{ik},\tau_{ik}/2)} \notag
\end{align}
in the non-sparse case for every $d\in A_{w_i}(u_{ij})$.
While in the sparse case, from
\begin{equation}
P(d_{ij}=d,\xi_{ij}=0|\dots) \propto \, \pi_i \mathbb{I}(u_{ij}<\tilde{w}_{id}), \, \quad d\in A_{\tilde{w}}(u_{ij}) \notag 
\end{equation}
where $A_{\tilde{w}}(u_{ij})=\{k: u_{ij}<\tilde{w}_k\}=\{0\}$, because $\tilde{w}_k=0,$ $\forall k>0$, 
\begin{align}
P(d_{ij}=d,\xi_{ij}=0|\dots)& \propto 
\left\{
\begin{array}{ll}
\pi_i \mathbb{I}(u_{ij}<1) \mathcal{N}(\beta_{ij}|0,\lambda_{ij}) \mathcal{G}a(\lambda_{ij}|\gamma_0,\tau_0/2)& \mathrm{if} \; d=0, \\
0& \mathrm{if} \; d>0.
\end{array}
\right. \notag 
\end{align}
\subsubsection*{Update the prior restriction probabilities \texorpdfstring{$\pi$}{pi}}
The full conditional for $\pi_i$ is,
\begin{equation}
f(\pi_i|\dots) \propto \, \mathcal{B}e\left( n_i+ 1 -\sum_{i=1}^{n_i} \mathbb{I}(\xi_{ii}=1), \alpha_i  + \sum_{i=1}^{n_i} \mathbb{I}(\xi_{ii}=1) \right). \notag
\end{equation}

\end{document}